\documentclass[12pt,preprint,flushrt]{emulateapj}
\usepackage{graphicx}
\usepackage{tabulary}
\usepackage{epsfig}
\usepackage{amssymb}
\usepackage{multirow}
\usepackage{appendix}
\usepackage{natbib}

\usepackage[pdftitle={Circumbinary Habitability Niches}, colorlinks =
  true, breaklinks = true, citecolor = blue, linkcolor = blue,
  urlcolor = blue, pdfauthor = {Collaborators}]{hyperref}

\newcommand{\BHMcalc}{{\tt BHMcalc }}
\newcommand{\beq}[1]{\begin{equation}\label{#1}}
\newcommand{\eeq}{\end{equation}}

\shorttitle{Circumbinary Niches}
\shortauthors{Mason, Zuluaga, Cuartas \& Clark}

\begin{document}
\title{Circumbinary Habitability Niches}

\author{Paul A. Mason\altaffilmark{1,3}, Jorge
  I. Zuluaga\altaffilmark{2}, Pablo
  A. Cuartas-Restrepo\altaffilmark{2}, Joni M. Clark\altaffilmark{3}}

\altaffiltext{1}{Department of Physics, University of Texas at El
  Paso, El Paso, TX 79968, USA}

\altaffiltext{2}{FACom - Instituto de F\'{\i}sica - FCEN, Universidad
  de Antioquia, Calle 70 No. 52-21, Medell\'{\i}n, Colombia}

\altaffiltext{3}{Department of Mathematics and Physical Sciences, New
  Mexico State University - DACC, Las Cruces, NM, 88003, USA}

%%%%%%%%%%%%%%%%%%%%%%%%%%%%%%%%%%%%%%%%%%%%%%%%%%%%%%%%%%%%%%%%%%%%%%%%%%%%%%%%%
%ABSTRACT
%%%%%%%%%%%%%%%%%%%%%%%%%%%%%%%%%%%%%%%%%%%%%%%%%%%%%%%%%%%%%%%%%%%%%%%%%%%%%%%%%

\begin{abstract}

Binaries could provide the best niches for life in the galaxy.  Though
counterintuitive, this assertion follows directly from stellar tidal
interaction theory and the evolution of lower mass stars.  There is
strong evidence that chromospheric activity of rapidly rotating young
stars may be high enough to cause mass loss from atmospheres of
potentially habitable planets.  The removal of atmospheric water is
most critical.  Tidal breaking in binaries could help reduce magnetic
dynamo action and thereby chromospheric activity in favor of life. We
call this the Binary Habitability Mechanism (BHM), that we suggest
allows for water retention at levels comparable to or better than
Earth.  We discuss novel advantages that life may exploit, in these
cases, and suggest that life may even thrive on some circumbinary
planets. We find that while many binaries do not benefit from BHM,
high quality niches do exist for various combinations of stars between
0.55 and 1.0 solar masses. For a given pair of stellar masses, BHM
operates only for certain combinations of period and eccentricity.
Binaries having a solar-type primary seem to be quite well suited
niches having wide and distant habitable zones with plentiful water
and sufficient light for photosynthetic life. We speculate that, as a
direct result of BHM, conditions may be suitable for life on several
planets and possibly even moons of giant planets orbiting some
binaries.  Lower mass combinations, while more restrictive in
parameter space, provide niches lasting many billions of years and are
rich suppliers of photosynthetic photons.  We provide a publicly
available web-site
(\href{http://bit.ly/BHM-calculator}{http://bit.ly/BHM-calculator},
\href{http://bit.ly/BHM-calculator-mirror}{http://bit.ly/BHM-calculator-mirror}),
which calculates the BHM effects presented in this paper.

\end{abstract}

%%%%%%%%%%%%%%%%%%%%%%%%%%%%%%%%%%%%%%%%%%%%%%%%%%%%%%%%%%%%%%%%%%%%%%%%%%%%%%%%%
%KEY WORDS
%%%%%%%%%%%%%%%%%%%%%%%%%%%%%%%%%%%%%%%%%%%%%%%%%%%%%%%%%%%%%%%%%%%%%%%%%%%%%%%%%


%%%%%%%%%%%%%%%%%%%%%%%%%%%%%%%%%%%%%%%%%%%%%%%%%%%%%%%%%%%%%%%%%%%%%%%%%%%%%%%%%
%PAPER CONTENT
%%%%%%%%%%%%%%%%%%%%%%%%%%%%%%%%%%%%%%%%%%%%%%%%%%%%%%%%%%%%%%%%%%%%%%%%%%%%%%%%%

%%%%%%%%%%%%%%%%%%%%%%%%%%%%%%%%%%%%%%%%%%%%%%%%%%%%%%%%%%%%%%%%%%%%%%%%%
\section{Introduction}
\label{sec:introduction}
%%%%%%%%%%%%%%%%%%%%%%%%%%%%%%%%%%%%%%%%%%%%%%%%%%%%%%%%%%%%%%%%%%%%%%%%%

The existence of life on planets within multiple star systems has been
long thought to be, at best, capable of providing only marginal
habitats. We recently introduced a mechanism \citep{Mason13} that we
will now refer to as the {\it Binary Habitability Mechanism} (BHM), by
which tidal interaction between stars in moderately close binaries
provide a means for improved habitability conditions.  Namely, for
some combinations of eccentricity and orbital period, a sufficiently
strong tidal torque exists between the stars. This torque, i.e. a
breaking tide, often results in the slowing of stellar rotation rates
to synchronous or pseudo-synchronous locking with the binary
period. The larger star synchronizes first and in many cases this
locking occurs on a time-scale faster than standard mass-loss torque
experienced by single stars. This rotational evolution reduces the
stellar magnetic dynamo thereby substantially reducing XUV flux and
stellar wind flux from coronal emission. Atmospheric erosion is
especially important before planets develop dynamos and therefore
have magnetic protection (see e.g. \citealt{Zuluaga13}).

The idea that planets in multiple star systems might be habitable is
not new \citep{Huang60}.  \citet{Harrington77} used numerical
integrations to show that planetary orbits may be stable in either
p-type (circumbinary) or s-type (satellite) configurations. In the
current paper, we focus on p-type planets as they may undergo tidal
evolution. Harrington found that stable orbits exist as long as one of
the semi-major axes is 3-4 times longer than the other. This result
was confirmed by \citet{Holman99}, providing the stability limitations
used in this paper.  \citet{Haghighipour09} addressed theoretical pros
and cons associated with the formation of planets in both p-type and
s-type planetary systems. The dynamical stability of exoplanets and
protoplanetary disks of binaries in the solar neighborhood were
studied numerically by \citet{Liu12}. Theoretical circumbinary ice
lines, assumed to separate rocky planet zones from giant planet zones
around stars, have been calculated by \citet{Clanton13}, who finds in
particular, that solar mass twins may have ice lines outside of the
stability limit as long as the binary separation is less than 1
AU. The theoretical binaries presented in the current study (see Table
\ref{tab:Niches}), are well within this limit, so they
potentially have rocky planets.

Several authors have developed prescriptions for determining the
locations and widths of habitable zones (HZs) surrounding binary
systems \citep{Mason12, Quarles12, Mason13, Kane13, Haghighipour13}.
The standard approach is to use the models of \citet{Kasting93},
updated by \citet{Kopparapu13} and derived for single stars, along
with various methods for dealing, most often approximately, with the
spectral characteristics of the stars.  Most recently, HZ definitions
around single main sequence stars have been extended by the inclusion
of a range of planetary masses \citep{Kopparapu14}.

Consequences for planetary habitability are not limited to the
restrictions placed by the standard HZ definition.  Exposure of
planetary atmospheres to high levels of XUV flux, like those orbiting
in the HZ of highly active stars, appear to have dramatic
consequences.

For single stars, relations exist between age, rotation rate, and
magnetic activity \citep{Basri87,Wood05}. Rapidly rotating stars are
luminous XUV sources, due to intense chromospheric activity, and
thereby undergo significant mass-loss \citep{Wood05}. What we like to
call ``stellar aggression'' poses high risks to weakly magnetized
planets. The effect lessens as the star loses angular momentum, via
mass loss.  These relationships have been used to evaluate the
evolution of stellar aggression and its role in terrestrial planet
habitability \citep{Griebmeier07, Zuluaga13}.

When these relationships are applied to binaries \citep{Mason13}, we
find that early tidal spin-down of one or both stars produces an
effective stellar rotational aging. Thus a reduction of stellar
aggression likely results in a reduction of mass-loss from planetary
atmospheres. However, there are two sides to the tidal interaction
coin. For some binary configurations, especially those with short
periods, $P_{bin} <$ 10 days, the stellar rotation effect increases
activity. Hence, XUV radiation photoionizes H and other atoms as well
as heats and expands the outer atmosphere. Thereby driving mass loss
from the atmosphere of exoplanets and thus reducing their chances for
habitability.

We begin, in section 2, by illustrating BHM calculations for an
ensemble of 12 p-type planetary systems. Different mass ratios of the binary are
explored with several periods and eccentricities. The HZ limits of
\citet{Kopparapu14} are applied and stellar models are used to
determine continuous habitable zone (CHZ) limits.  In each case, a
hypothetical Earth-like planet is placed at the inner edge of the
CHZ. Stellar rotational evolution and the resulting XUV and SW fluxes
as well as planetary atmosphere mass loss rates are calculated as a
function of time. Integrated XUV and stellar wind values are compared
to Earth conditions. We find niches comparable to or in some cases
superior to that enjoyed by Earth. We present four examples in
detail. Specifically niches N1, N6, N7, and N9, are discussed in order
to illustrate advantages some circumbinary planets have over single
star planetary systems.

In section 3, we identify several factors promoting habitability in
BHM protected planetary systems. These include 1) increased water
retention (allowing lower mass planets), 2) multiple habitable circumbinary planets
(due to wide and distant CHZs in some cases),
 3) extended habitability lifetimes (for lower mass primaries), and 4)
high photosynthetic photon flux density (for photosynthetic biomass production).

Finally, in section 4, we summarize our results and suggest that BHM
candidates should be among the targets selected for observation in
searches for Earth like planets such as the Transiting Exoplanet
Survey Satellite (\textit{TESS}) \citep{Ricker10} and the Planetary Transits 
and Oscillations of stars telescope (\textit{PLATO})
\citep{Rauer13}.

\bigskip 
\bigskip

%%%%%%%%%%%%%%%%%%%%%%%%%%%%%%%%%%%%%%%%%%%%%%%%%%%%%%%%%%%%%%%%%%%%%%%%%
\section{The Binary Habitability Mechanism}
\label{sec:BHM}
%%%%%%%%%%%%%%%%%%%%%%%%%%%%%%%%%%%%%%%%%%%%%%%%%%%%%%%%%%%%%%%%%%%%%%%%%

In moderately close binaries, tidal interaction between stellar
components is able to erode rotation and ultimately leave both
stars in resonant states where the period of rotation is closely
related to the binary orbital period \citep{Mason13}.  Since rotation
is closely correlated to chromospheric activity, a star whose rotation
has been sufficiently reduced will be less aggressive, in terms of XUV
and stellar wind (hereafter SW) flux. Hence, there will be a
corresponding reduction in planetary atmospheric mass loss. Our
results suggest that planets will find better conditions to maintain
life in circumbinary HZ orbits, when the BHM operates, as compared to
single star systems.

In order to apply the mechanism, we model several aspects of the
binary system including: (1) the limits of the circumbinary habitable
zone as a function of the luminosity of the stars; (2) the continuous
circumbinary habitable zone, by incorporating stellar evolution models; 
(3) the star-star tidal interaction resulting in the rotational breaking of
both stars; and (4) the binary extrapolation of single star
relationships between stellar rotational period, magnetic activity,
XUV, and SW evolution in order to derive (5) planetary atmospheric
mass-loss. An outline of BHM calculations and preliminary results for six Kepler
circumbinary planets were presented by \citet{Mason13}.

In order to provide the reader an opportunity to explore BHM parameter
space we have developed and tested a publicly available on-line tool,
the ``BHM Calculator'', \BHMcalc
(\href{http://bit.ly/BHM-calculator}{http://bit.ly/BHM-calculator}).
A mirror site of the calculator is also available
(\href{http://bit.ly/BHM-calculator-mirror}{http://bit.ly/BHM-calculator-mirror}).
The spirit of this tool is similar to that created by
\citet{Muller14}.  However, our aim is not only to provide renditions
of the instantaneous circumbinary HZ, but also to calculate BHM
properties of the system including those related to the rotational
evolution of the stellar components and the combined XUV and SW fluxes
as measured at different distances from the binary.  Moreover, our
tool provides numerical results that can be further manipulated and
used to calculate other properties. For details of the models used by
the \BHMcalc and their validation please refer to \citet {Mason13}.

%TTTTTTTTTTTTTTTTTTTTTTTTTTTTTTTTTTTTTTTTTTTTTTTTTTTTTTTTTTTTTTTTTTTTTTTTTTTTTTTT
%TABLE
%TTTTTTTTTTTTTTTTTTTTTTTTTTTTTTTTTTTTTTTTTTTTTTTTTTTTTTTTTTTTTTTTTTTTTTTTTTTTTTTT
\begin{table*}[ht]
\tiny
\centering 
\begin{tabular}{ccccccccccc}
  \hline\hline 

  Niche & $M_{1}/M_\odot$ & $M_{2}/M_\odot$ & q & $P_{\rm bin}$
  & $e_{\rm bin}$ & $a_{\rm crit}$ & CHZ$_{\rm in}$ - CHZ$_{\rm out}$ & CHZ width &
  $\eta_{\rm XUV}$ & $\eta_{\rm SW}$ \\ 

  \null & \null & \null & \null & [days] & \null & [AU] & [AU] & [AU] & 
  BHM/no BHM & BHM/no BHM \\ \hline

  \multicolumn{10}{c}{Solar-Mass Primary}\\\hline
  N1, \href{http://bit.ly/BHM-niche1}{http://bit.ly/BHM-niche1} & 1.00 & 1.00 & 1.00 & 15 & 0.1 & 0.15 & 1.57 - 2.18 & 0.61 & 0.562/0.998 & 0.229/0.998 \\ 
  N2, \href{http://bit.ly/BHM-niche2}{http://bit.ly/BHM-niche2} & \null & 0.85 & 0.85 & 20 & 0.5 & 0.65 &  1.25 - 1.86 & 0.61 & 0.973/1.497 & 0.343/1.423 \\ 
  N3, \href{http://bit.ly/BHM-niche3}{http://bit.ly/BHM-niche3} & \null & 1.00 & 1.00 & 40 & 0.3 & 0.92 & 1.57 - 2.18 & 0.61 & 0.799/0.992 & 0.273/0.992 \\ 
  N4, \href{http://bit.ly/BHM-niche4}{http://bit.ly/BHM-niche4} & \null & 0.70 & 0.70 & 15 & 0.4 & 0.50 & 1.15 - 1.68 & 0.53 & 1.106/1.576 & 0.371/1.524 \\ 
  N5, \href{http://bit.ly/BHM-niche5}{http://bit.ly/BHM-niche5} & \null & 1.00 & 1.00 & 60 & 0.5 & 1.36 & 1.57 - 2.18 & 0.61 & 0.803/0.986 & 0.273/0.986 \\ 
  N6, \href{http://bit.ly/BHM-niche6}{http://bit.ly/BHM-niche6} & \null & 0.55 & 0.55 & 12 & 0.1 & 0.33 & 1.12 - 1.60 & 0.48 & 1.013/1.480 & 0.376/1.465 \\ 
  \hline\multicolumn{10}{c}{$M_{\star}=0.85$ Primary}\\\hline
  N7, \href{http://bit.ly/BHM-niche7}{http://bit.ly/BHM-niche7} & 0.85 & 0.85 & 1.00 & 12 & 0.2 & 0.36 & 0.89 - 1.56 & 0.67 & 0.508/0.919 & 0.223/0.919 \\ 
  N8, \href{http://bit.ly/BHM-niche8}{http://bit.ly/BHM-niche8} & \null & 0.70 & 0.82 & 20 & 0.3 & 0.54 & 0.71 - 1.32 & 0.61 & 1.007/1.263 & 0.362/1.278 \\ 
  N9, \href{http://bit.ly/BHM-niche9}{http://bit.ly/BHM-niche9} & \null & 0.55 & 0.65 & 15 & 0.3 & 0.44 & 0.66 - 1.20 & 0.54 & 1.010/1.318 & 0.388/1.356 \\ 
  \hline\multicolumn{10}{c}{$M_{\star}=0.70$ Primary}\\\hline
  N10, \href{http://bit.ly/BHM-niche10}{http://bit.ly/BHM-niche10} & 0.70 & 0.70 & 1.00 & 18 & 0.4 & 0.51 & \textbf{0.51} - 1.06 & 0.55 & 0.759/0.889 & 0.256/0.889 \\ 
  N11, \href{http://bit.ly/BHM-niche11}{http://bit.ly/BHM-niche11} & \null & 0.55 & 0.79 & 30 & 0.4 & 0.71 & \textbf{0.71} - 0.90 & 0.19 & 0.375/0.398 & 0.127/0.408 \\ 
  \hline\multicolumn{10}{c}{$M_{\star}=0.55$ Twins}\\\hline
  N12, \href{http://bit.ly/BHM-niche12}{http://bit.ly/BHM-niche12} & 0.55 & 0.55 & 1.00 & 20 & 0.3 & 0.47 & \textbf{0.47} - 0.73 & 0.26 & 0.360/0.382 & 0.124/0.381  \\ \hline
\end{tabular}
\vspace{2mm}
\caption{Selected binary habitability niches. Here the masses of the
  primary and secondary are $M_{1}$ and $M_{2}$ respectively,
  $q=M_{2}/M_{1}$, $e_{\rm bin}$ and $P_{\rm bin}$ are the
  eccentricity and period of the binary system, $a_{\rm crit}$ is the
  critical distance for orbital stability, CHZ$_{\rm in}$ and
  CHZ$_{\rm out}$ are the inner and outer edges of the circumbinary
  continuous habitable zone and $\eta_{\rm XUV,SW}$ are the asymptotic
  time-integrated XUV and SW fluxes (see Eq. \ref{eq:eta}). We compare
  the integrated fluxes when the BHM is operating and when we assume
  it is not.  In cases for which the calculated CHZ$_{\rm in}$ is
  greater than the critical limit (N10, N11 and N12), the CHZ$_{\rm
    in}$ is set to be equal to the critical stability limit (values in
  boldface).  We have provided links to the results page of the
  \BHMcalc for each Niche.
  \label{tab:Niches}}
\end{table*}
%TTTTTTTTTTTTTTTTTTTTTTTTTTTTTTTTTTTTTTTTTTTTTTTTTTTTTTTTTTTTTTTTTTTTTTTTTTTTTTTT

For illustration purposes, we examine an ensemble of 12 binary
configurations derived from four different primary masses, 1.00, 0.85,
0.70 and 0.55 $M_{\odot}$ (see Table \ref{tab:Niches}). Here and 
elsewhere in this paper, we define the masses of the primary and 
secondary as $M_{1}$ and $M_{2}$ respectively, $a_{bin}$ is the 
semi major axis of the binary, $e_{bin}$ is the binary eccentricity, and 
$P_{bin}$ is the orbital period of the binary. The semi-major axis 
of the planet is a, and the age of the system is $\tau$ . 

The configurations shown in Table \ref{tab:Niches} are selected 
from among a continuum of binary configurations
for which we have found more favorable habitability conditions. The
BHM calculator generates stellar rotational evolution, time evolution
and integrated XUV flux, time evolution and integrated SW flux, 
integrated planetary atmospheric mass-loss, and insolation and 
photosynthetic photon flux density plots.  We evaluate
integrated XUV and SW proxies for Earth-like habitability in order to
quantify stellar aggression.  When XUV and SW proxies are 
less than corresponding values for Earth, we call these favorable 
configurations, {\it Binary Habitability Niches}.

For each potential niche, we compute the limits of the
circumbinary continuous habitable zone (CHZ), defined as the region
where habitable conditions, in terms of insolation and following the
criteria defined by \citet{Kopparapu14}, are maintained during the
main sequence phase of the primary component.  The edges of the
circumbinary ``instantaneous'' HZ have been estimated using 
refinements of the model proposed in \citep{Mason13}.  
We have verified that our estimations are consistent with
independent results obtained by  \citet{Kane13}, \citet{Haghighipour13},
and  \citet{Muller14}.  For calculating HZs we have used the extreme
limit criteria given as ``Recent Venus'' (RV) for the inner edge and
``Early Mars'' (EM) for the outer edge of the HZ.  The evolution of
stellar properties have been obtained from a fine grid of theoretical
isochrones calculated by the Padova Group \citep{Girardi00}.
%CHECK LAST SENTENCE (PADOVA)

%FFFFFFFFFFFFFFFFFFFFFFFFFFFFFFFFFFFFFFFFFFFFFFFFFFFFFFFFFFFFFFFFFFFFFFFFFFFFFFFF
%FIGURE 1
%FFFFFFFFFFFFFFFFFFFFFFFFFFFFFFFFFFFFFFFFFFFFFFFFFFFFFFFFFFFFFFFFFFFFFFFFFFFFFFFF
\begin{figure*}
\centering
\vspace{0.2cm}
\includegraphics[width=0.60\textwidth]{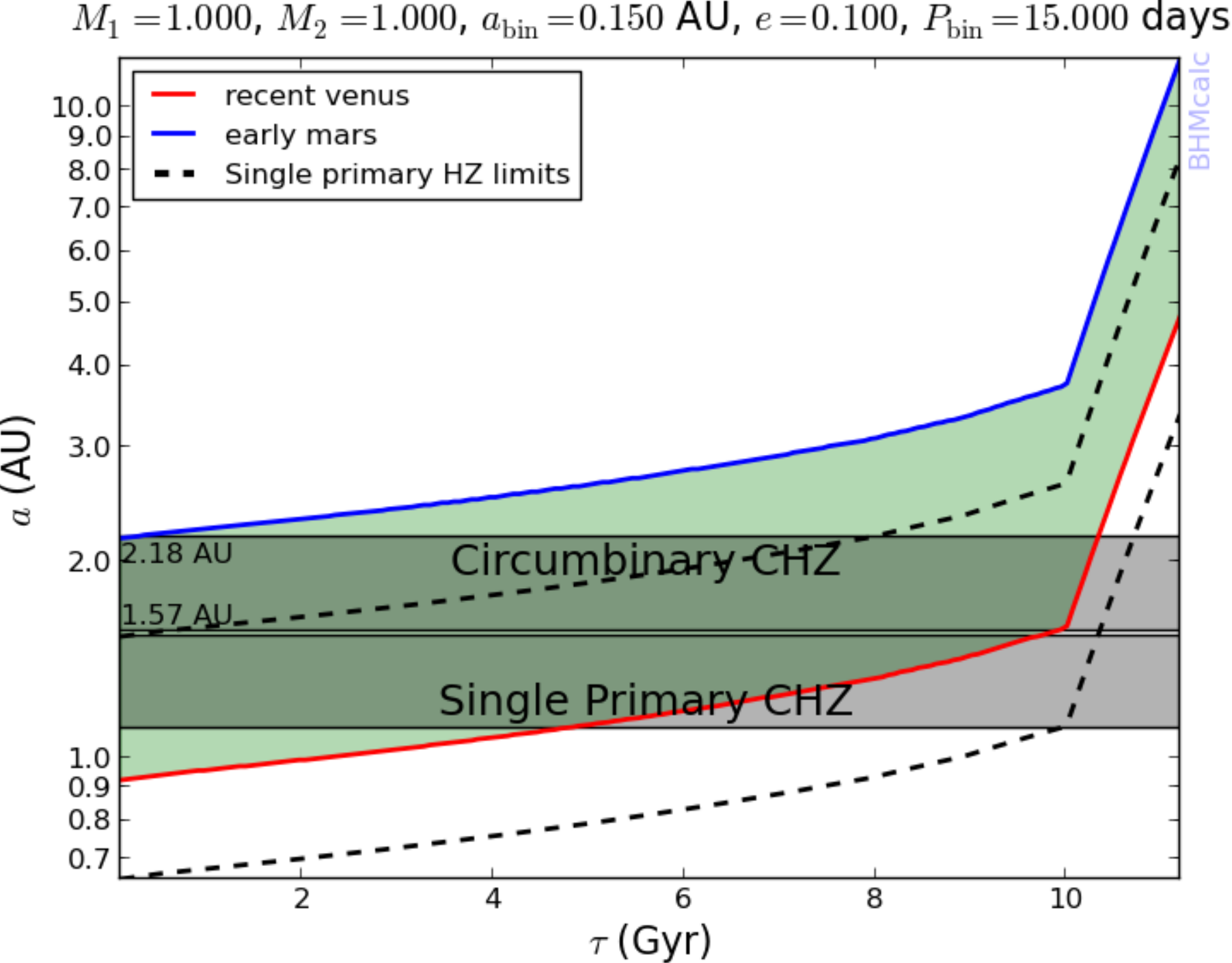}
\caption{Evolution of the habitable zone around a single solar-mass
  star (dashed lines) and solar twins (continuous lines) of niche N1. 
  The masses of the primary and secondary are $M_{1}$
  and $M_{2}$ respectively, $a_{bin}$ is the semimajor axis of the binary,
   e is the binary eccentricity, and $P_{bin}$ is the orbital period of the binary.
  The semi-major axis of the planet, a, is plotted as a function of
   the age of the system, $\tau$. The stars are considered to be 
  coevol. At $\tau$=10 Gyrs, the primary star abandons the 
  main sequence.  The position of the inner edge of the HZ at this time 
  defines the innermost limit of the continuous habitable zone (CHZ). 
  Since the vertical scale is logarithmic, the width of the single star CHZ and
  the circumbinary CHZ are shaded and appear equal, but do not overlap.  
  However, the circumbinary CHZ
  ranges from CHZ$_{\rm in}=$1.57 AU to CHZ$_{\rm in}=$2.18 AU, while
  the single CHZ goes from 1.11 AU to 1.53 AU, i.e. the circumbinary
  CHZ of N1 is 45\% wider and 42\% more distant than in the single star
  case.}
\label{fig:SolarCHZ}
\end{figure*}
%FFFFFFFFFFFFFFFFFFFFFFFFFFFFFFFFFFFFFFFFFFFFFFFFFFFFFFFFFFFFFFFFFFFFFFFFFFFFFFFF

%FFFFFFFFFFFFFFFFFFFFFFFFFFFFFFFFFFFFFFFFFFFFFFFFFFFFFFFFFFFFFFFFFFFFFFFFFFFFFFFF
%FIGURE 2
%FFFFFFFFFFFFFFFFFFFFFFFFFFFFFFFFFFFFFFFFFFFFFFFFFFFFFFFFFFFFFFFFFFFFFFFFFFFFFFFF
\begin{figure*}
\centering
\includegraphics[width=0.49\textwidth] {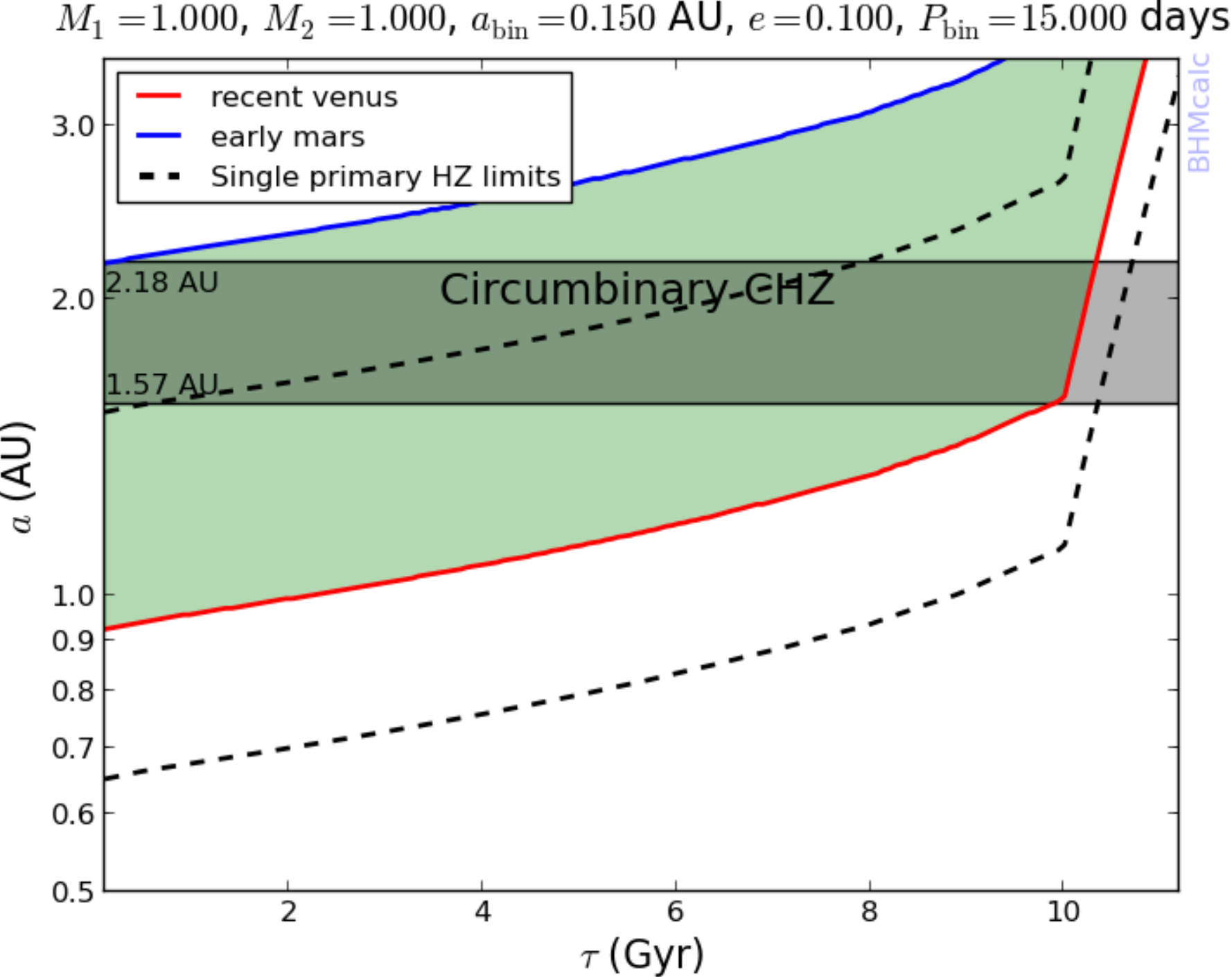}
\includegraphics[width=0.49\textwidth] {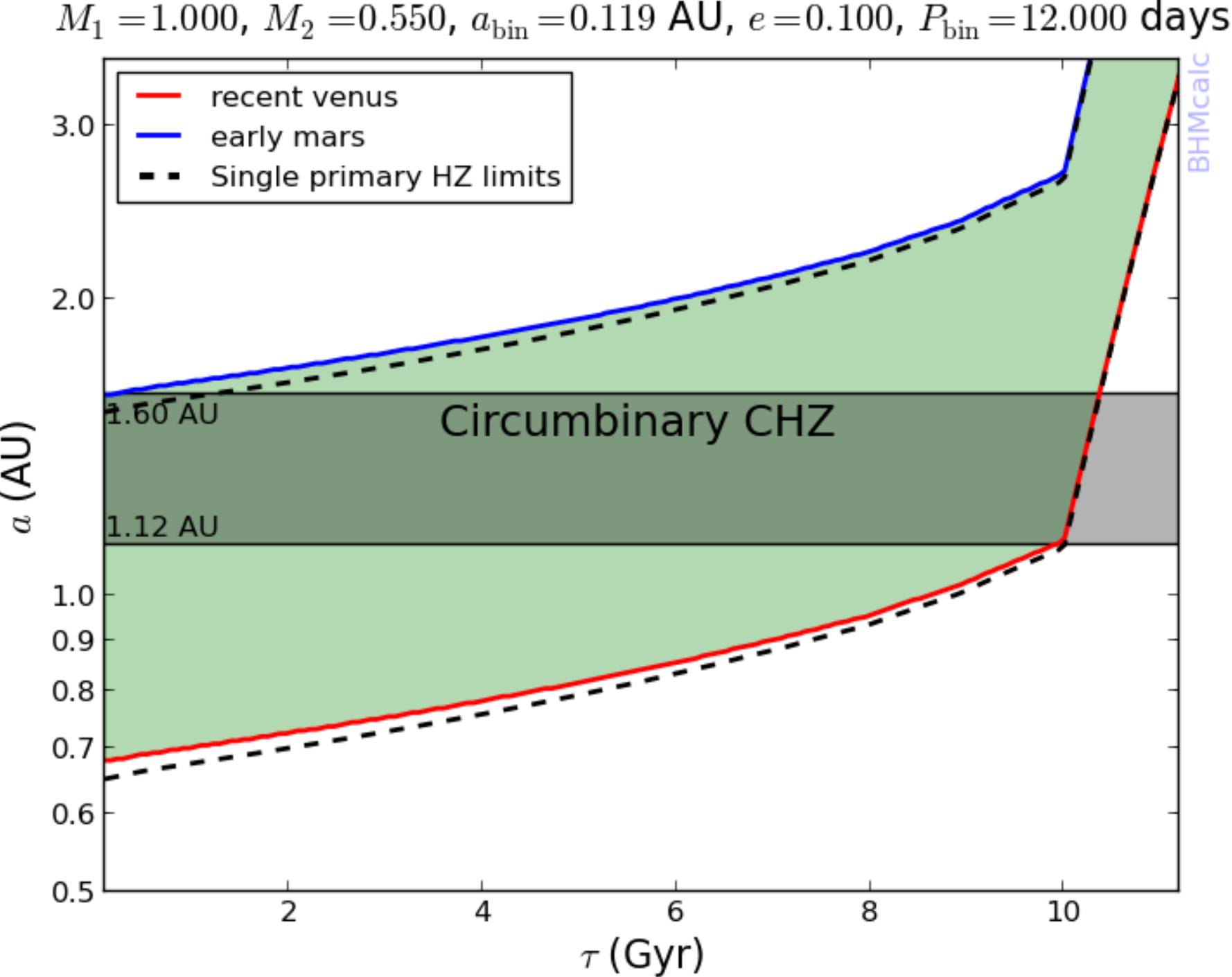}
\includegraphics[width=0.49\textwidth] {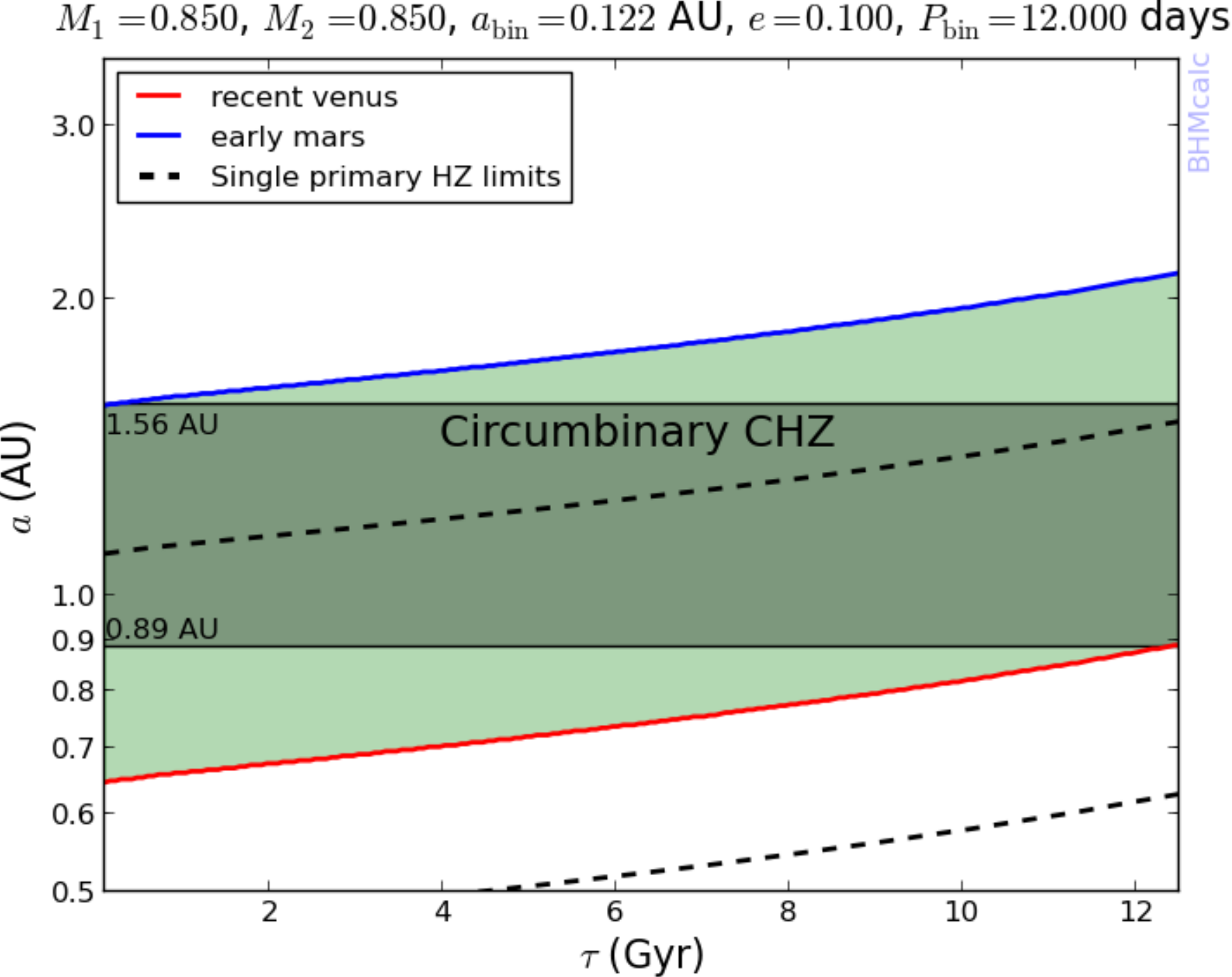}
\includegraphics[width=0.49\textwidth] {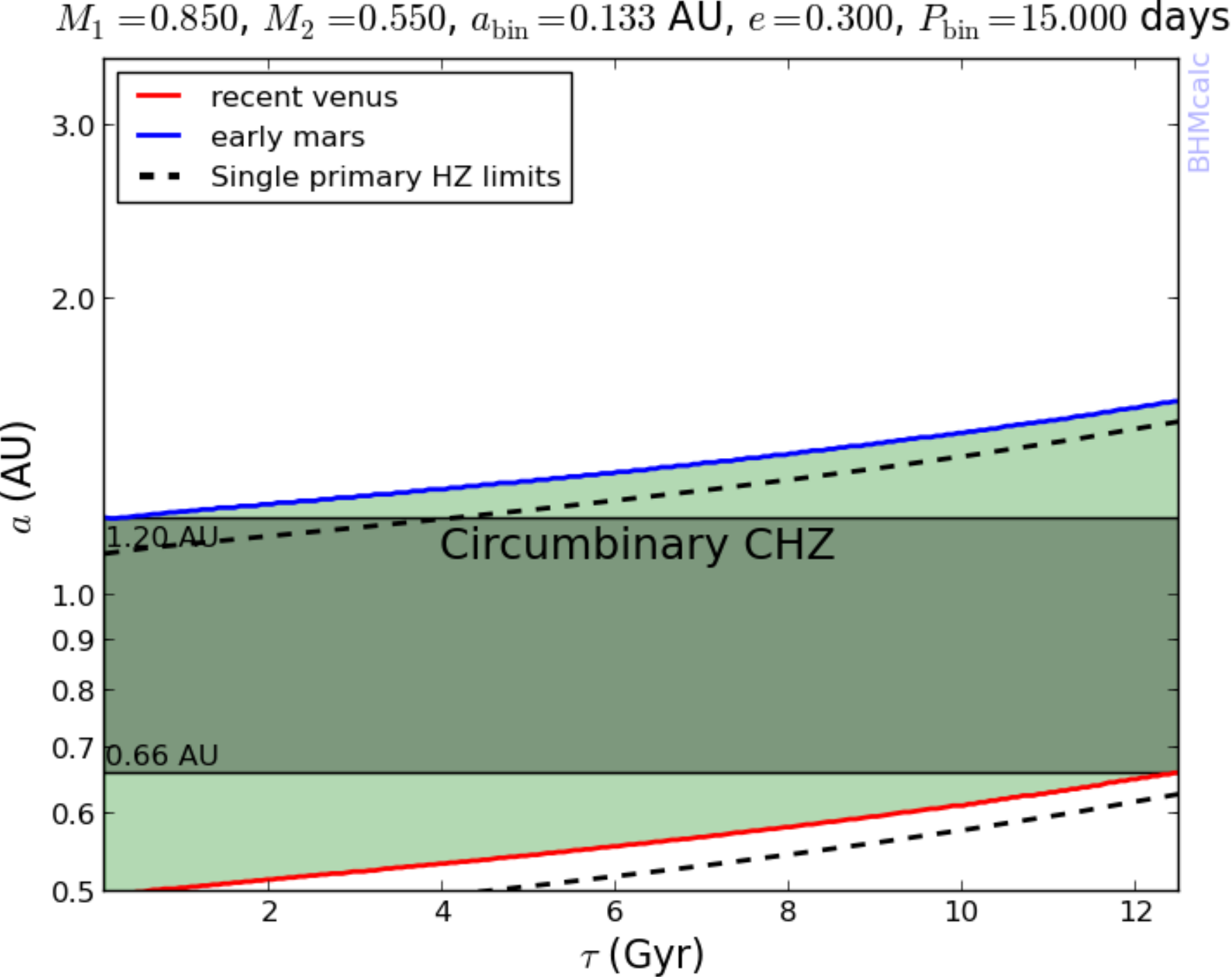}
\caption{The CHZ for selected binary systems, N1, N6, N7, and N9 are shown. 
  The vertical axes are all shown on the same scale for easy comparison. The top two
  panels (niches N1 and N6) contain solar mass primaries. For maximum
  longevity, those planets should reside at distances that are greater than
  the middle of the continuous habitable zone (shaded region). On the
  other hand, the bottom two panels (niches N7 and N9),
   involve lower mass primaries with planets optimally 
  positioned at the inner edge of the continuous habitable zone. 
  The horizontal scales and the corresponding CHZ limits for lower 
  mass primaries are based on a final age of 12.5 Gyr, based on the age of the 
  galaxy. See text for more details.
\label{fig:CHZ-Niches}}
\end{figure*}
%FFFFFFFFFFFFFFFFFFFFFFFFFFFFFFFFFFFFFFFFFFFFFFFFFFFFFFFFFFFFFFFFFFFFFFFFFFFFFFFF

To illustrate differences between single and circumbinary CHZs we
present Figure \ref{fig:SolarCHZ}
showing the evolution of the
instantaneous circumstellar and circumbinary HZ limits for a single
solar-mass star and for solar-twins (niche N1 in Table
\ref{tab:Niches}).  The limits of the CHZ in both cases are
highlighted. In order to compare CHZs for niches N1, N6, N7, and N9,
the CHZ for these cases are shown in Figure \ref{fig:CHZ-Niches}.

In order to evaluate how favorable a potential binary is, we need to
calculate the level of stellar aggression to which a planet at a given
distance from the binary is subject, compared with the level of
aggression experienced by Earth.  We have identified two proxies for
aggression: (1) the XUV flux, i.e. the combined flux in extreme
ultraviolet and in X-rays and (2) the SW flux.  High levels of
XUV flux are responsible for enhanced non-thermal and thermal
mass-loss from planetary atmospheres (see e.g. \citealt{Tian09}).
In addition, high SW fluxes are potentially able to remove tens
to hundreds of bars from unmagnetized Earth-like planets
\citep{Zendejas10,Lammer09c,Lammer12}.

\begin{figure*}
\centering
\includegraphics[width=0.49\textwidth] {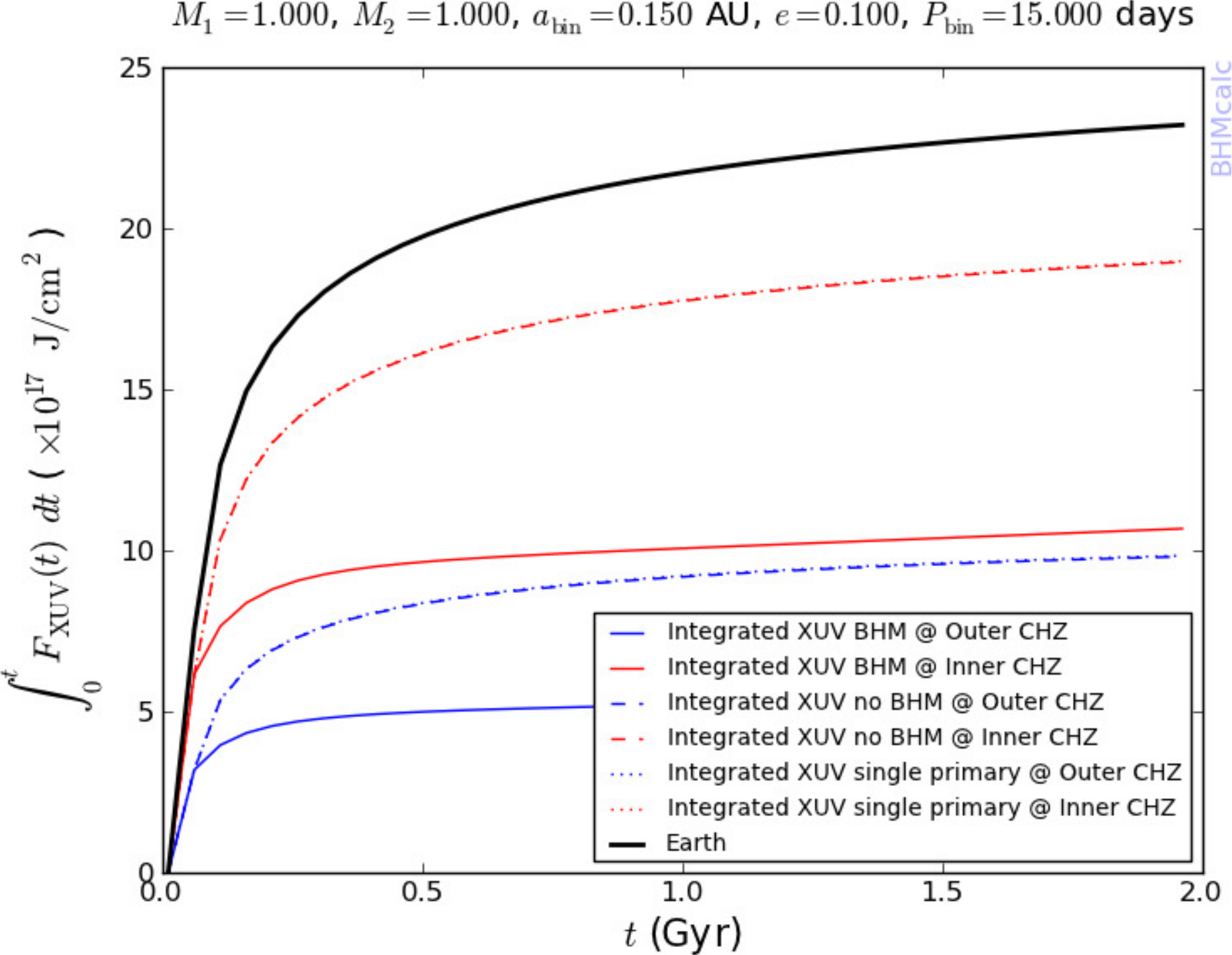}
\includegraphics[width=0.49\textwidth] {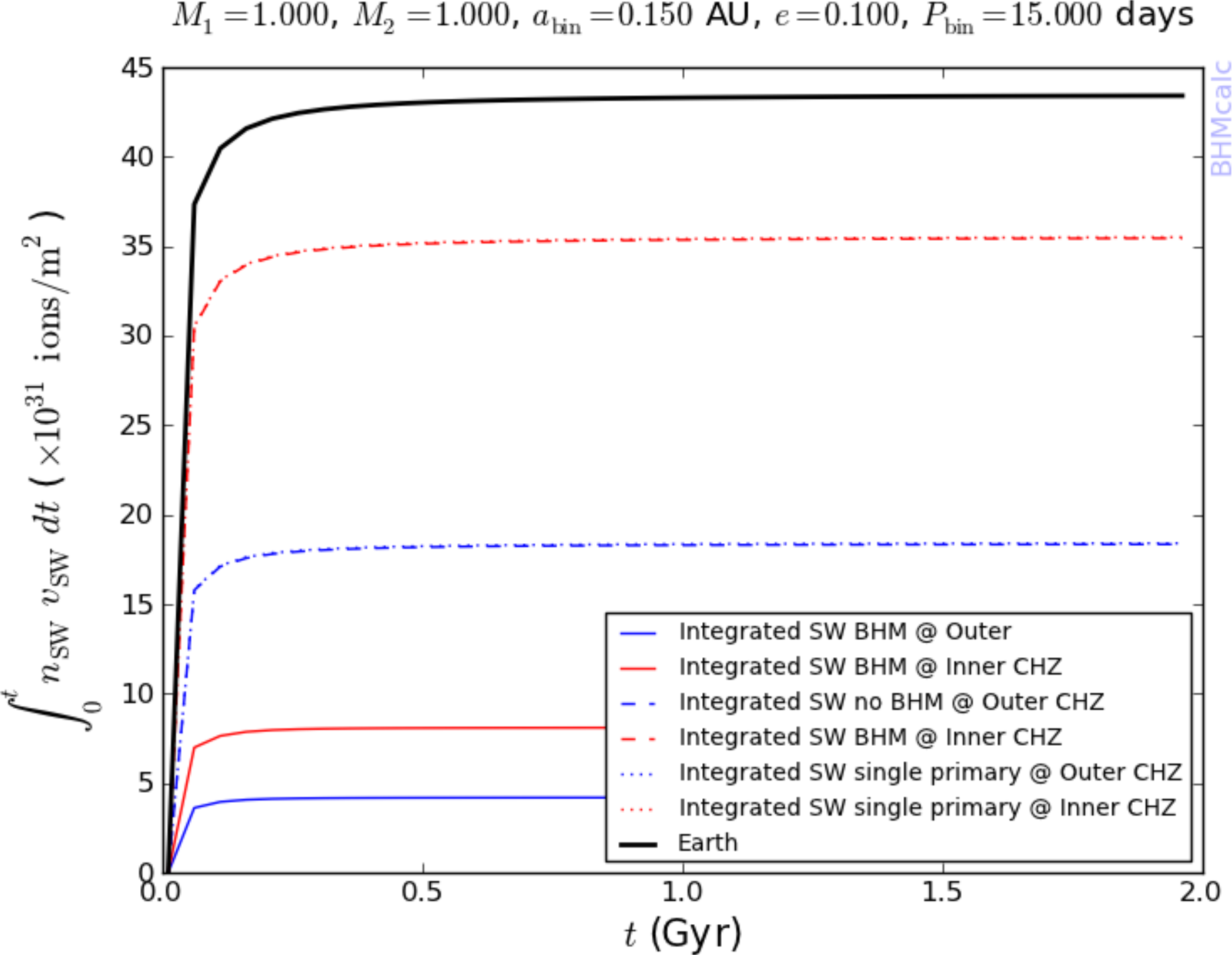}
\caption{Time-integrated XUV (left) and SW (right) fluxes on the CHZ of solar twins
  (Niche 1), shown including BHM (solid lines) and without BHM (dashed
  lines). For comparison purposes we have included the same quantity
  calculated in the CHZ of a solar-mass single star (single primary,
  dotted lines) and more specifically at the Earth distance (thick
  solid line). Note that in both cases the inner CHZ with no BHM is slightly 
  below Earth values, this is because the Earth is closer to the sun than the
  solar inner CHZ, which is at 1.1 AU.
  \label{fig:Integrated}}
\end{figure*}

Since most of the aggression occurs during the earliest phases of
stellar evolution, the time-integrated XUV and SW fluxes reach an
asymptotic value within the first Gyr or so. Figure
\ref{fig:Integrated} shows the results of such integrations in the
case of a solar-mass single star and solar twins at different
positions inside their respective CHZs.

%FFFFFFFFFFFFFFFFFFFFFFFFFFFFFFFFFFFFFFFFFFFFFFFFFFFFFFFFFFFFFFFFFFFFFFFFFFFFFFFF
%FIGURE 3
%FFFFFFFFFFFFFFFFFFFFFFFFFFFFFFFFFFFFFFFFFFFFFFFFFFFFFFFFFFFFFFFFFFFFFFFFFFFFFFFF
%FFFFFFFFFFFFFFFFFFFFFFFFFFFFFFFFFFFFFFFFFFFFFFFFFFFFFFFFFFFFFFFFFFFFFFFFFFFFFFFF

In order to compare to Earth level aggression, we have also calculated
the asymptotic time-integrated fluxes on an Earth-like planet, with
semi major axis a, located at the inner edge of the CHZ of a solar-mass 
star ($a$=1.11 AU) and at
the analogous position, but around each binary habitability niche.  
The ratio of the latter to the former, $\eta_{\rm Y}$, is:
%CHECK

\beq{eq:eta}
\eta_{\rm Y}=
\frac{\int_0^{\tau} F^{\rm bin}_{\rm Y}(t,a={\rm CHZ}_{\rm in})\,dt}
{\int_0^{\tau} F^{\rm sing}_{\rm Y}(t,a={\rm CHZ}^{\rm sing}_{\rm in})\,dt}
\eeq

where $\tau$ is an arbitrary time larger than 1 Gyr (we have assumed
$\tau$=2 Gyr) and $Y$ stands for either XUV or SW.  

In Table \ref{tab:Niches}, input parameters for the selected
binary habitability niches are provided as well as model outputs for
CHZ and stellar aggression proxies.  Notice that for mass ratios close
to 1, as well as large eccentricities and periods, the critical
distance $a_{\rm crit}$ \citep{Mason13} can be inside of the habitable
zone as in examples N5, N10, N11, and N12, which have been
highlighted in bold text.  In these cases, the inner CHZ values are determined by
the orbital stability limit rather than the HZ limit.

The examples in Table \ref{tab:Niches} are selected in order to cover a
range of mass ratios as well as several equal mass pairs and a range
of orbital parameters. These are not chosen to be ideal or
comprehensive in any way, only to provide examples for illustration.

For brevity, we focus the rest of the discussion on four examples, 
N1, N6, N7, and N9.
Evolution of the HZ for two cases involving
equal mass stars and two cases with disparate stars are shown.
Niche N1, solar mass twins, and niche N7, 0.85
$M_{\odot}$ twins, are shown in the left panels of Figure
\ref{fig:CHZ-Niches}.  Right panels of the same figure show examples
of lower mass ratio binaries. Specifically, niche N6, a solar mass primary with a
0.55 $M_{\odot}$ companion, and niche N9, a 0.85 $M_{\odot}$ primary
with a 0.55 $M_{\odot}$ secondary are shown. Two important properties 
are apparent. Binaries with a lower mass primary, (bottom two panels)
have a wider CHZ, due to the larger increase in luminosity of solar mass
stars (top two panels) as a function of time. Also, twins (left two panels) 
generally have wider CHZs than disparate binaries (right two panels) due
to the minimal flux variation of twins as a function of binary phase.

Binaries with differing mass components and short periods or small
eccentricities have critical distances well inside the inner habitable
zone limit. The ability for a planet to remain in the CHZ for the
entirety of the primary star's evolutionary lifetime, requires it to
be in the outer region of the CHZ near the beginning. For these
purposes we define the optimal distance, for solar or greater mass
primaries, as being the one that places the planet in the HZ for the
longest period of time; i.e. the early-Mars limit at ZAMS.   
Those binaries containing a solar mass primary have long-lived
niches for planets residing at distances that are greater than the
middle of the CHZ. However, there is a trade-off.  Distant planets
receive less radiation for photosynthetic consumption than closer
ones.  So, for binaries with lower mass primaries, and hence long
lifetimes; the optimum planetary distance is near the inner edge 
of the CHZ. This position maximizes
stellar insolation as long as stellar aggression is sufficiently low.

\begin{figure*}
\centering
\includegraphics[width=0.49\textwidth] {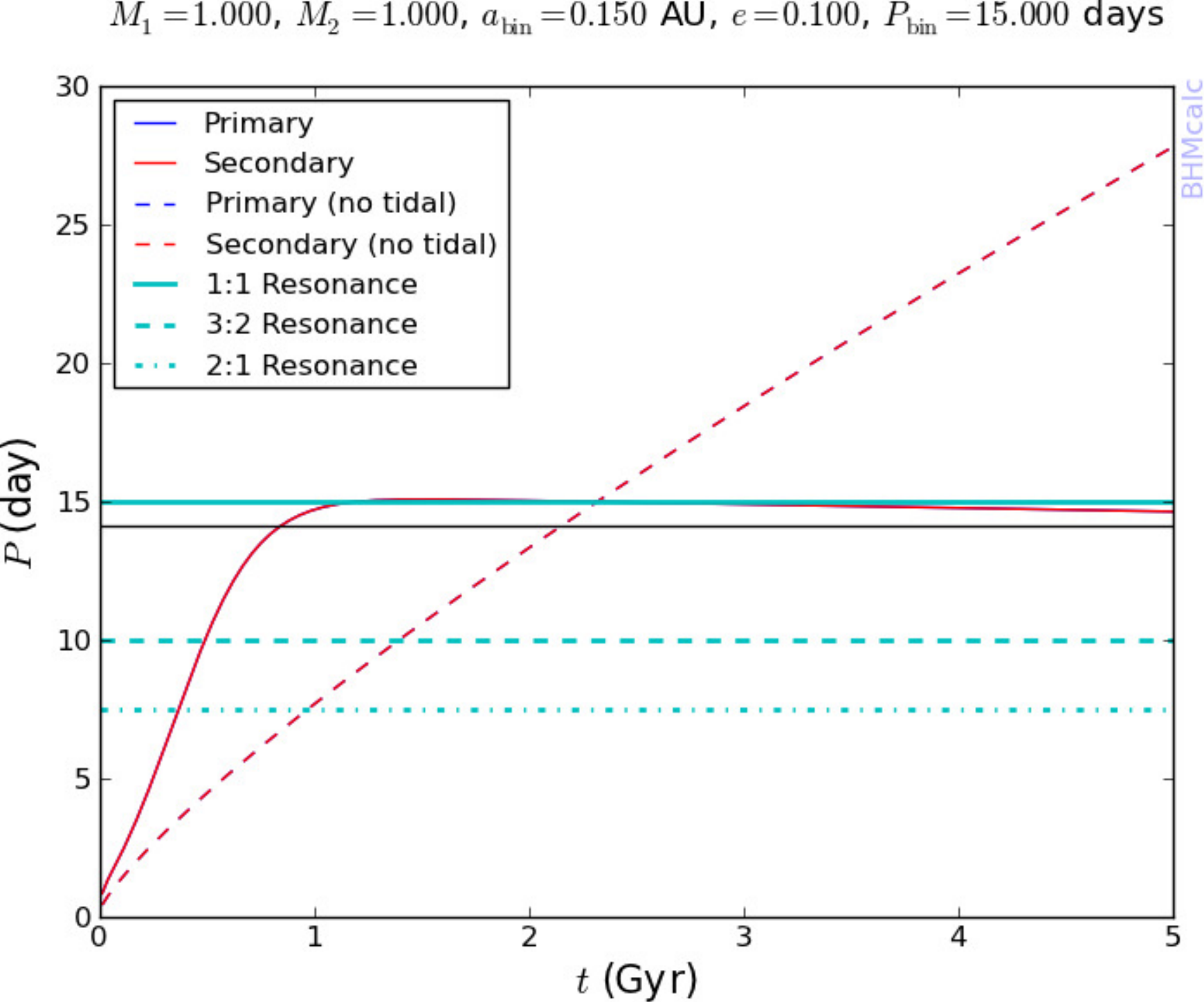}
\includegraphics[width=0.49\textwidth] {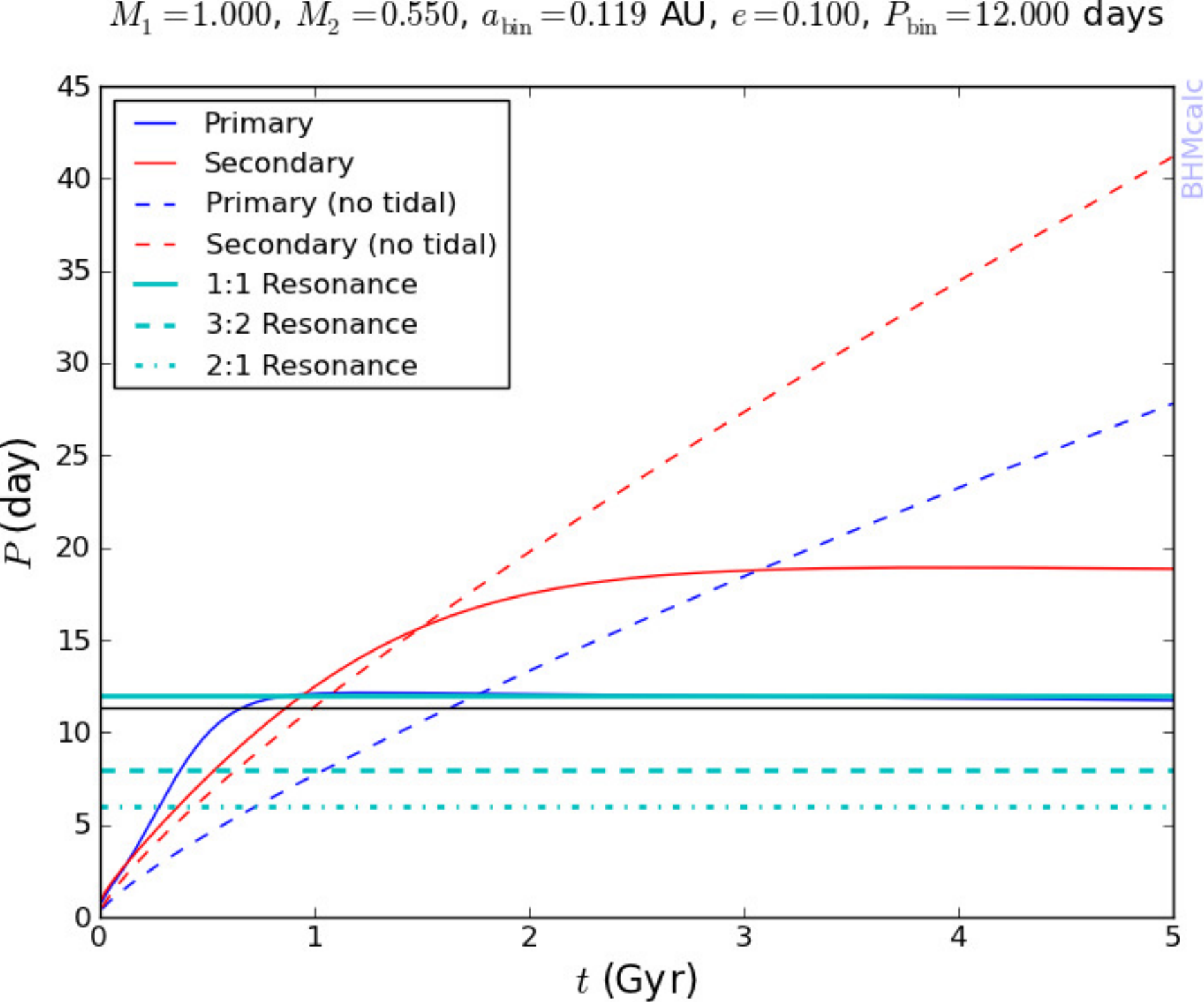}
\includegraphics[width=0.49\textwidth] {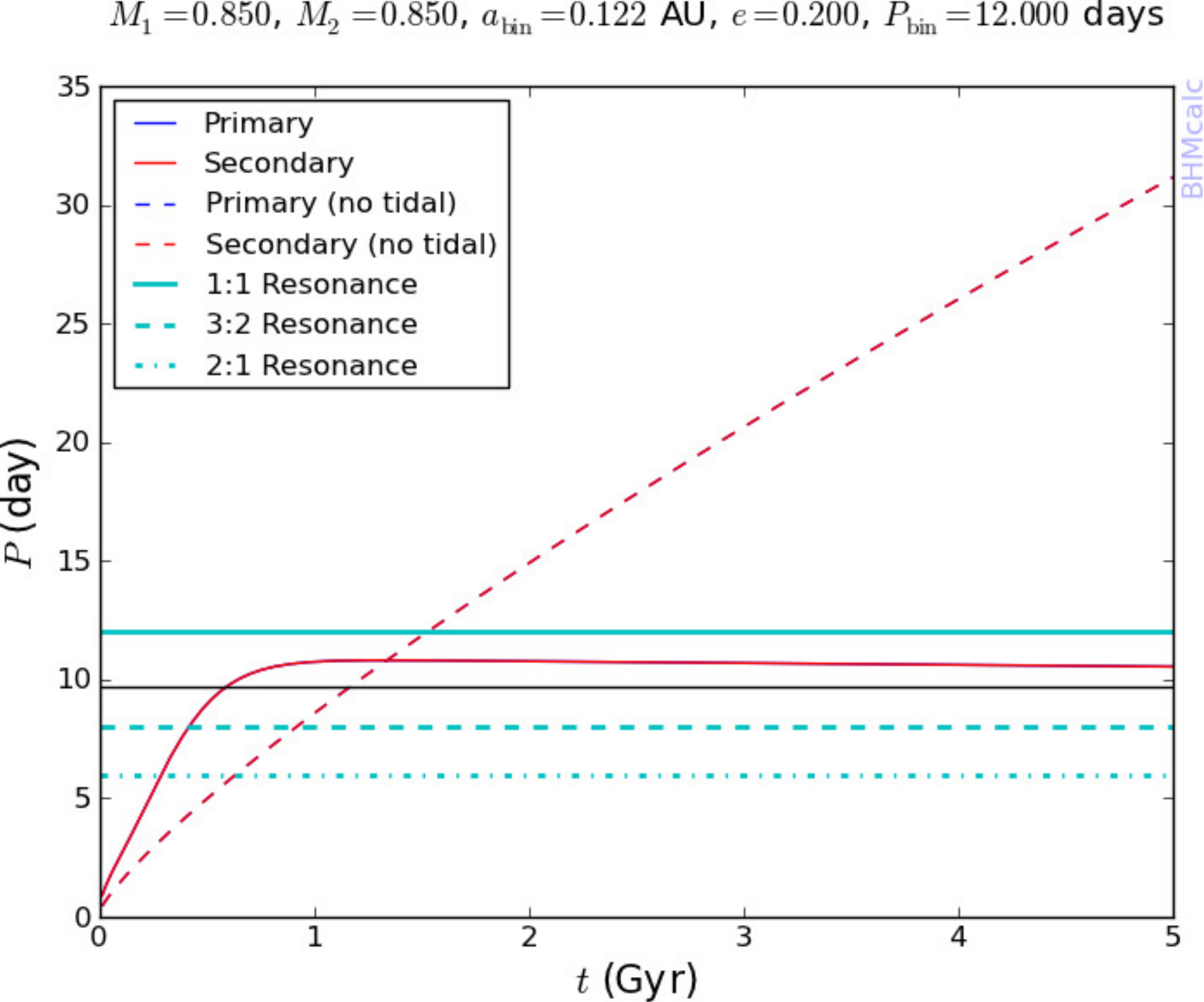}
\includegraphics[width=0.49\textwidth] {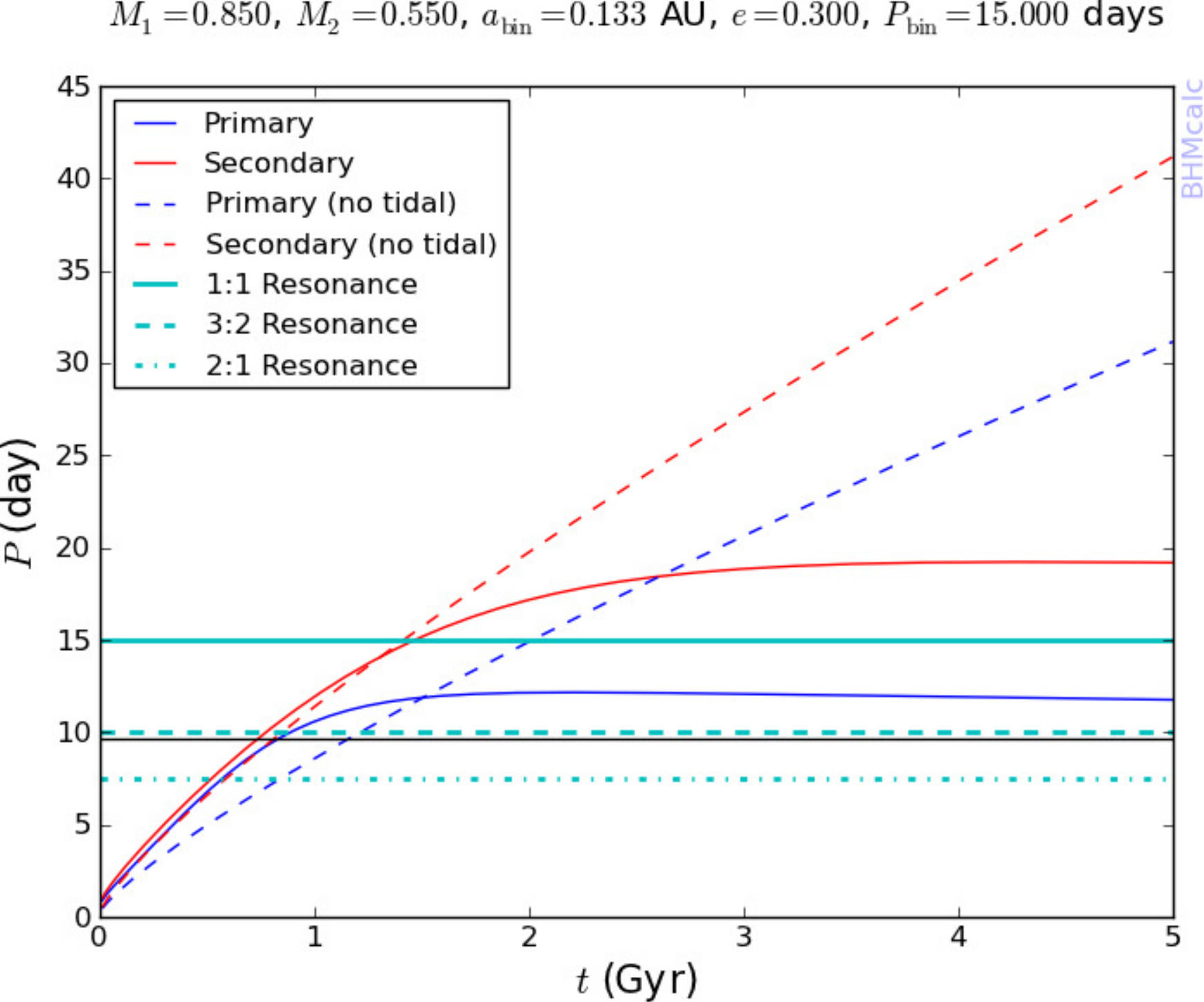}
\caption{Stellar rotation period evolution for niches N1, N6, N7, and
  N9; the same examples as those shown in Figure 2. 
  These models include both the tidal synchronization torque of
  the binary and the standard single star mass loss torque. The left
  two panels show 1.0 and 0.85 $M_\odot$ twins, so the stellar
  components evolve to the same period at the same time. The right
  panels show two different mass ratios and one can see that the
  primary synchronizes first and in cases with higher eccentricity,
  such as N9, in the bottom right panel, the primary
  pseudo-synchronizes well short of the binary period while the
  secondary synchronizes at a rotational period that is few days
  longer than the binary period.
 \label{fig:Pevol}}
\end{figure*}

The essence of the BHM effect is the early tidal evolution of stellar
rotation in moderately close binaries.  We model the tidal torque
using the formalism of \citet{Hut81} and \citet{Zahn08}. 
However, we include both the tidal synchronization
torque and the standard single star mass loss torque to determine the
time evolution of the rotation of both stars. Rotational evolution
results for niches N1, N6, N7, and N9 are shown in Figure
\ref{fig:Pevol}. This selection provides a illustration of the variety
of effects encountered. For example, in the top left plot involving
solar mass twins at low eccentricity, niche N1, both stars synchronize
to the 15 day binary period in 0.5 Gyr.  The lower left plot,
niche N7, however, shows synchronization of the primary in 1 Gyr, but
only a mild increase in stellar rotation period of the secondary until
2 Gyr.  Notice for non-zero eccentricity, such as the top right plot,
niche N6, a case of a solar mass primary and M-type companion with e =
0.1, the primary synchronizes to the binary period of 12
days. However, the rotation period of the secondary synchronizes over
a longer time and to a longer period.

%FFFFFFFFFFFFFFFFFFFFFFFFFFFFFFFFFFFFFFFFFFFFFFFFFFFFFFFFFFFFFFFFFFFFFFFFFFFFFFFF
%FIGURE 4
%FFFFFFFFFFFFFFFFFFFFFFFFFFFFFFFFFFFFFFFFFFFFFFFFFFFFFFFFFFFFFFFFFFFFFFFFFFFFFFFF
%FFFFFFFFFFFFFFFFFFFFFFFFFFFFFFFFFFFFFFFFFFFFFFFFFFFFFFFFFFFFFFFFFFFFFFFFFFFFFFFF

\begin{figure*}
\centering
\includegraphics[width=0.49\textwidth] {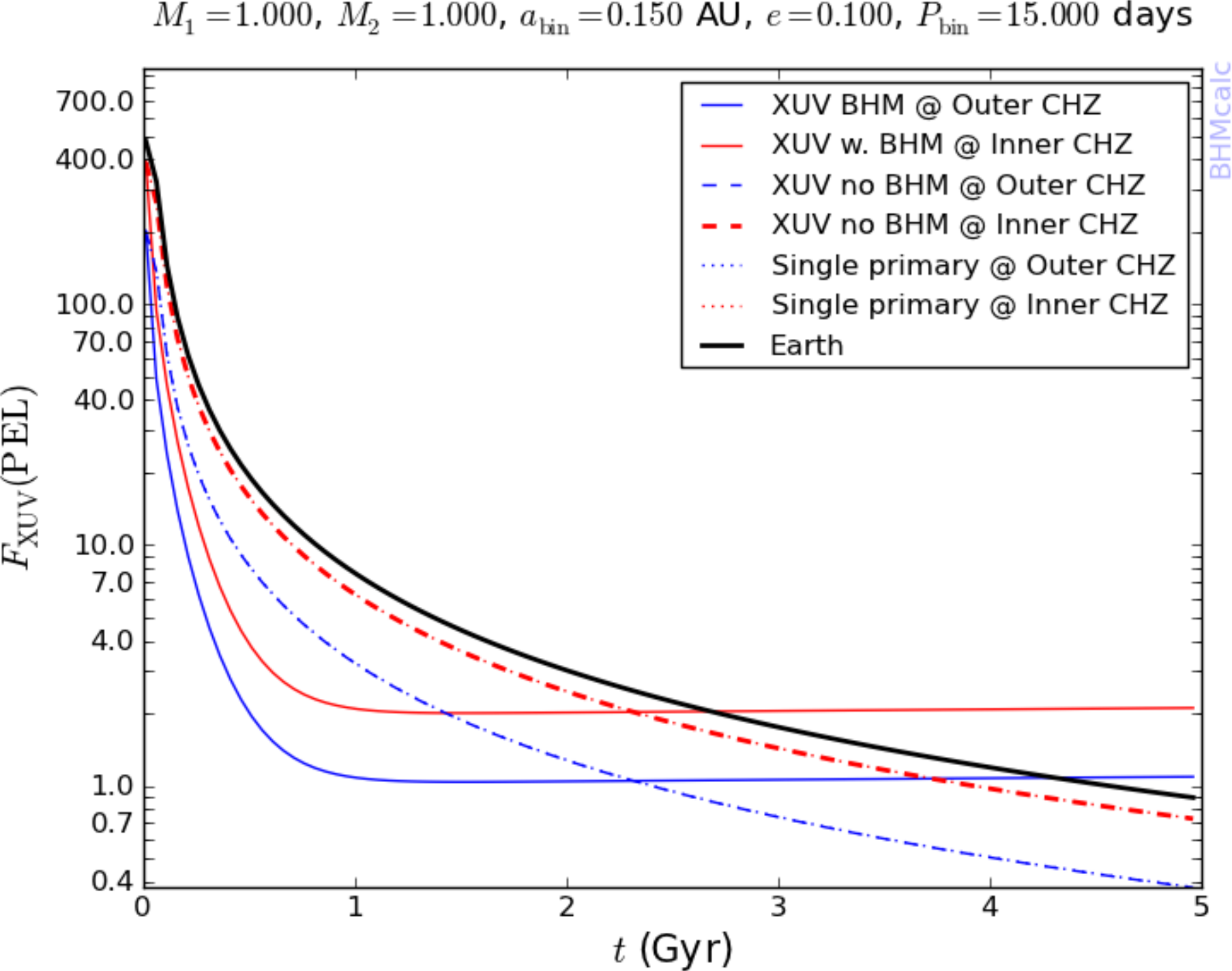}
\includegraphics[width=0.49\textwidth] {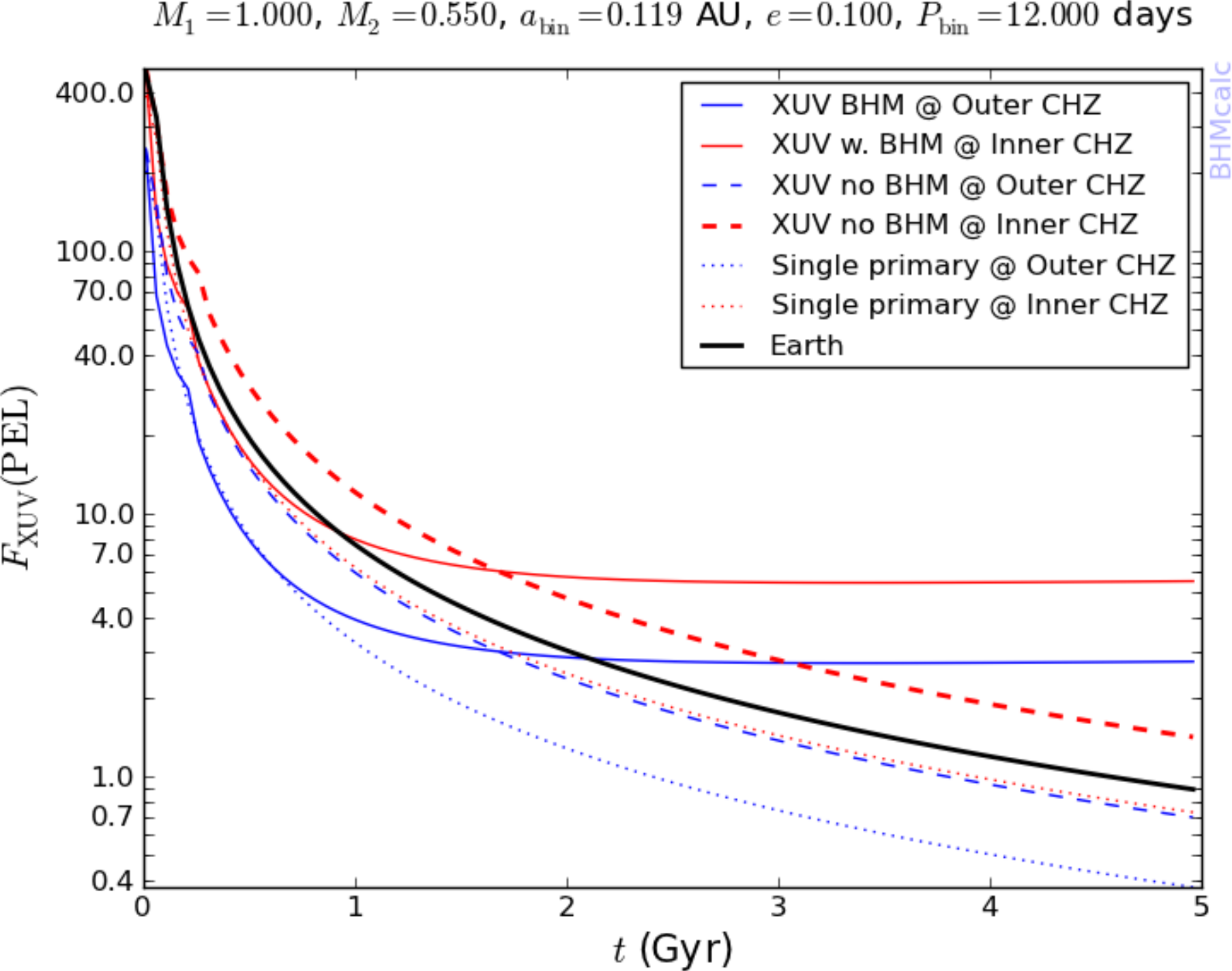}
\includegraphics[width=0.49\textwidth] {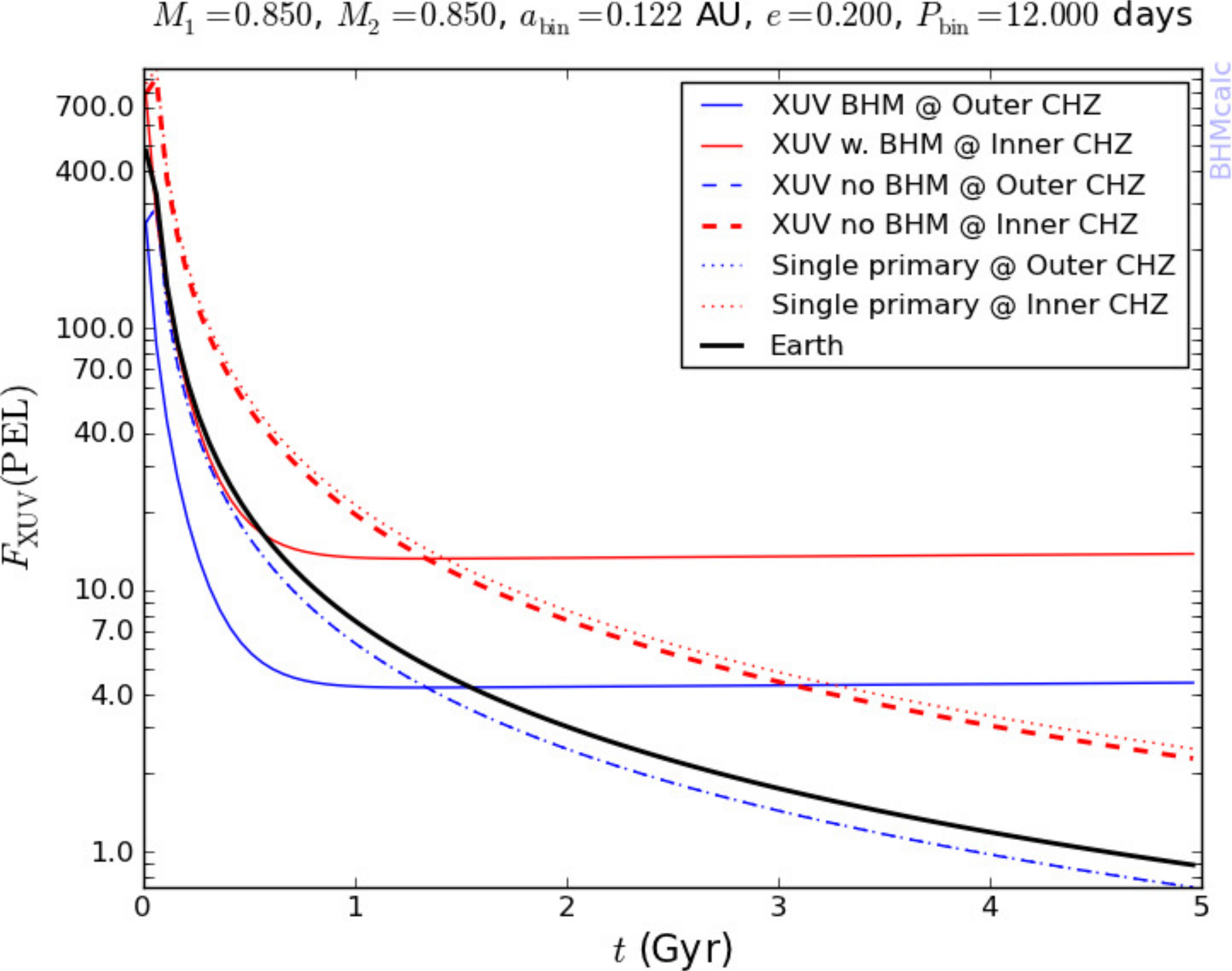}
\includegraphics[width=0.49\textwidth] {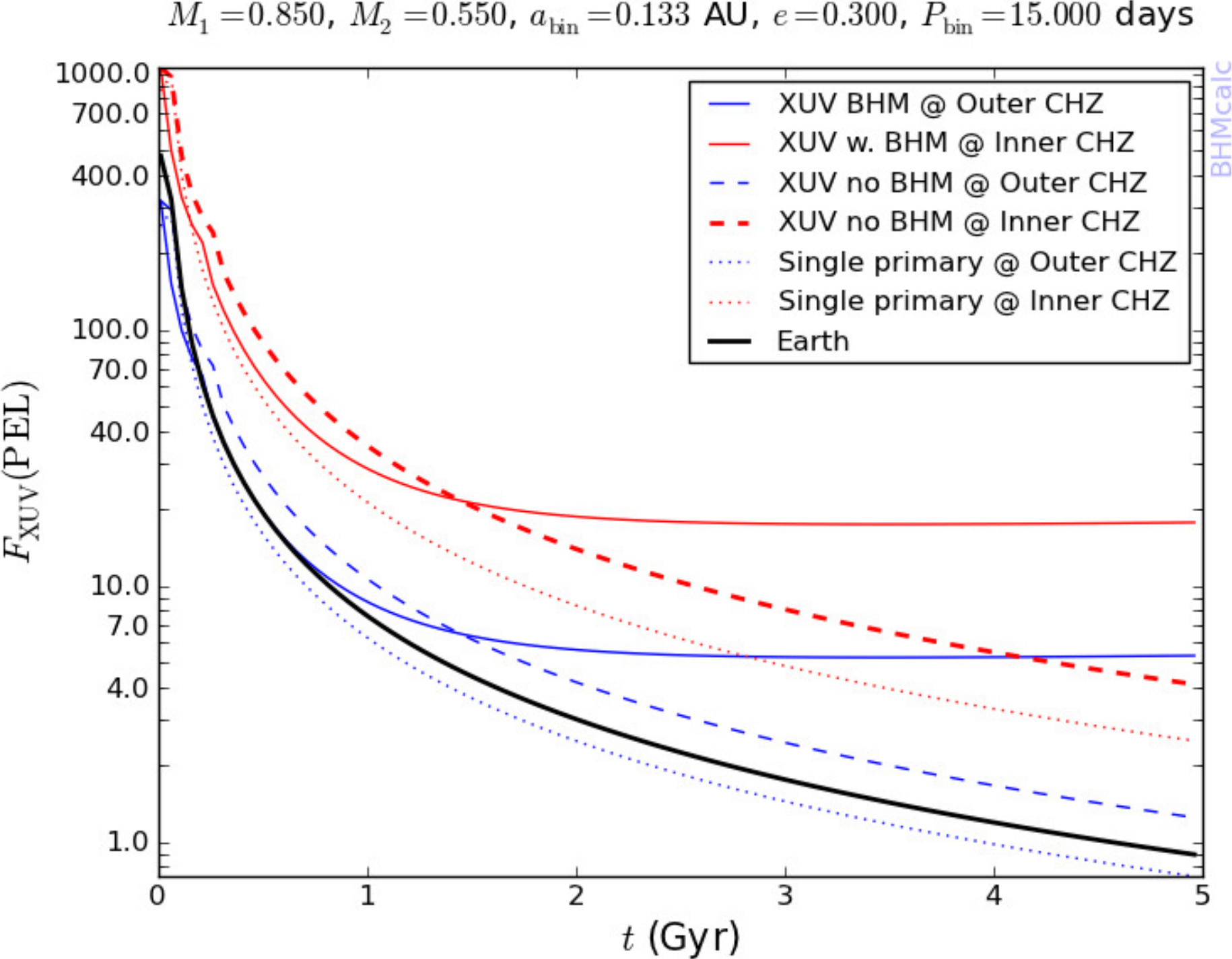}
\caption{Stellar XUV evolution for niches N1, N6, N7, and N9; 
the same examples as those shown in Figures 2 and 4. In each
  case, there is an early decrease in XUV flux.  
  The effect of strong tidal breaking (see Figure 1) on XUV emission during first Gyr is
  clear as XUV flux drops precipitously.
  Tidal breaking results in a total integrated XUV flux
  significantly less than that incident on Earth at all points within
  the CHZ (see Table \ref{tab:Niches}).
\label{fig:XUVevol}}
\end{figure*}

These early rotational evolution effects provide enhanced habitability
over single stars during the early aggressive stages of stellar
evolution. In Figure \ref{fig:XUVevol} the evolution XUV flux for the
same niches shown in Figures \ref{fig:CHZ-Niches} and \ref{fig:Pevol}
are presented. Time evolution plots show the dramatic decrease in
aggression during the first Gyr in these cases.  

%FFFFFFFFFFFFFFFFFFFFFFFFFFFFFFFFFFFFFFFFFFFFFFFFFFFFFFFFFFFFFFFFFFFFFFFFFFFFFFFF
%FIGURE 5
%FFFFFFFFFFFFFFFFFFFFFFFFFFFFFFFFFFFFFFFFFFFFFFFFFFFFFFFFFFFFFFFFFFFFFFFFFFFFFFFF
%FFFFFFFFFFFFFFFFFFFFFFFFFFFFFFFFFFFFFFFFFFFFFFFFFFFFFFFFFFFFFFFFFFFFFFFFFFFFFFFF

\begin{figure*}
\centering
\includegraphics[width=0.6\textwidth]{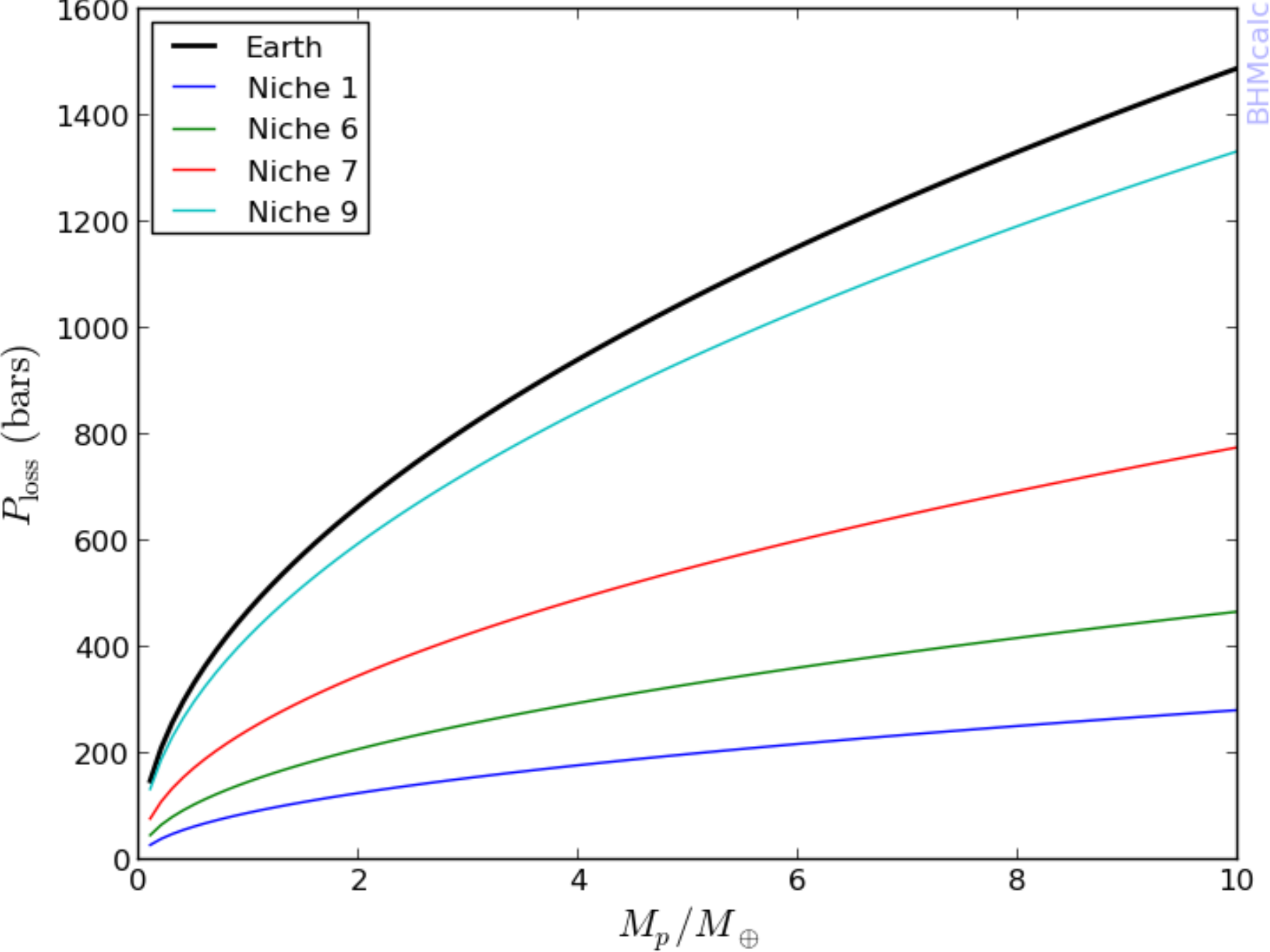}
\caption{Atmospheric mass-loss for circumbinary planets in niches
  N1, N6, N7, N9, (the same niches as those presented in Figures 2, 4, and 5) 
  at the inner edge of the CHZ are shown as a function of planetary mass.
  The corresponding estimate for the Earth is shown for reference. In each of these 
  cases, mass loss for an Earth like 
  planet is less than that experienced by Earth at all points within the 
  circumbinary HZ.\label{fig:MassLoss}\vspace{0.2cm}}
\end{figure*}

%THIS IS THE PARAGRAPH ON MASS-LOSS
In Figure \ref{fig:MassLoss}, the time integrated mass loss derived for 
niches N1, N6, N7, and N9, is shown as a function of planetary mass.
For each niche, we consider a planet located at the
inner edge of the CHZ, providing an upper limit.
Mass-loss is compared to planets located at 1 AU
from a single solar-mass star (i.e. at an Earth-like
location). Mass loss is calculated for a non-magnetized planet having an
atmosphere of arbitrary mass. That is, an atmosphere with no limits on the
amount of mass that could be stripped-off by the stellar wind.  For
this purpose, we use the simplified model of
\citet{Zendejas10}. We assume an entrapment factor of
$\alpha=0.3$ (see Eq. 2 in \citealt{Zendejas10}) and an atmosphere
composed primarily of CO$_2$.  Massive planets, for a given density,
lose more mass because
they have a larger atmospheric cross-section exposed to the erosion of the
stellar wind. Moreover, by expressing a given value of the total mass 
loss in bars it is clear that more massive planets, having larger gravities 
will also lose more pressure.
Significantly, mass-loss for any of the binary niches examined here
is less than planets in an Earth like location, for any planetary mass 
and any point within the CHZ.

%% Note, that unlike Figures \ref{fig:CHZ-Niches} and
%% \ref{fig:Pevol}; Figure \ref{fig:MassLoss} shows integrated values, so
%% as the curves flatten, mass loss decreases to near zero.  In most of
%% these cases, the integrated planetary atmospheric mass loss is less
%% than or comparable to that suffered by Earth, even at the inner edge
%% of the CHZ.  Notice from Table \ref{tab:Niches}, that atmospheric mass
%% loss is significantly greater than Earth values in niches N10, N11,
%% N12.  In those cases, a more massive planet than Earth might be
%% required to reach Earth mass loss levels. Likewise, for low mass loss
%% rates, like those shown, sub-Earth mass planets might undergo mass
%% loss similar to Earth levels.

%FFFFFFFFFFFFFFFFFFFFFFFFFFFFFFFFFFFFFFFFFFFFFFFFFFFFFFFFFFFFFFFFFFFFFFFFFFFFFFFF
%FIGURE 6
%FFFFFFFFFFFFFFFFFFFFFFFFFFFFFFFFFFFFFFFFFFFFFFFFFFFFFFFFFFFFFFFFFFFFFFFFFFFFFFFF
%FFFFFFFFFFFFFFFFFFFFFFFFFFFFFFFFFFFFFFFFFFFFFFFFFFFFFFFFFFFFFFFFFFFFFFFFFFFFFFFF
%%%%%%%%%%%%%%%%%%%%%%%%%%%%%%%%%%%%%%%%%%%%%%%%%%%%%%%%%%%%%%%%%%%%%%%%%

%%%%%%%%%%%%%%%%%%%%%%%%%%%%%%%%%%%%%%%%%%%%%%%%%%%%%%%%%%%%%%%%%%%%%%%%%
\section{Binary Benefits}
\label{sec:Benefits}
%%%%%%%%%%%%%%%%%%%%%%%%%%%%%%%%%%%%%%%%%%%%%%%%%%%%%%%%%%%%%%%%%%%%%%%%%

Rather than discussing the limits of habitability as a function of
atmospheric and geophysical variants, this discussion focuses on
specific examples of binaries that provide conditions that are likely
to promote planetary habitability equal to or beyond that experienced
by Earth. The idea that Earth may not represent ideal conditions for
the origin and maintenance of life, especially complex or intelligent
life, is proposed by \citet{Heller14}, who argue that Earth-like
habitability (in single star systems) may be enhanced by tweaking
factors, such as: (1) reducing stellar mass, thereby increasing the
lifetime of the habitable zone; (2) increasing planetary mass, thereby
increasing the strength and lifetime of the planet's magnetic
protection (see \citealt{Zuluaga13}) and (3) increasing the semi-major
axis of the planet, in order to increase its time in the habitable
zone; just to name a few. In this paper, we show that this concept of
super habitability naturally extends to circumbinary planets. So here,
we elaborate on some exceptional circumbinary planet configurations.
%FINAL

We identify four benefits that a minority of main sequence binaries
possess as the direct result of BHM operation.
These are:
%FINAL

\begin{enumerate}

  \item \textbf{Water retention}.  Binaries could provide
    opportunities for lower mass planets and those with weaker
    magnetic protection as well as inner HZ (Venus-like) planets to
    preserve its atmosphere and key volatiles and hence to have
    better chance for habitability.
    %FINAL

\bigskip

  \item \textbf{Multiplanet habitability}. With ample water around
    some binaries, BHM might make it possible to have more than one
    habitable planet. Many circumbinary CHZs are wider and more distant
    than the solar CHZ. This can be potentially an advantage for a hypothetical
    panspermia mechanism as well as other potential advantages for
    advanced life.
    %FINAL

  \item \textbf{Long habitability lifetimes}.  Binary systems with
    lower mass primaries could provide moist habitats for a time-scale
    longer than the Hubble time.  These binaries include stars that
    would likely be too active to sustain life as we know it as a
    single star.
    %FINAL

  \item \textbf{Photosynthetic photon flux density}. The amount of light 
  useful for photosynthesis is sufficient in many
    cases to drive photosynthesis and some circumbinary planets may
    have more photon flux density that that received by Earth,
    especially near the inner edge of the HZ.
    %FINAL

\end{enumerate}

Not all of these benefits are shared by all habitable circumbinary
planets. Some details on each of these follow.

\subsection{Water Retention}

If the period of a binary is longer than about 10 days and short
enough and/or eccentric enough to allow for the operation of BHM;
then synchronization of stellar rotation with binary rotation occurs.
Planets within the circumbinary habitable zone experience reduced
stellar aggression and potentially retain moist atmospheres.  Details
concerning the effects of increased, or decreased, XUV flux and
stellar winds on planetary atmospheres and life are complex. The
retention of water on the surface and in the atmosphere depends
sensitively on the atmospheric composition \citep{Tian09} as well as
geophysical factors such as volcanic out-gassing fluxes and
composition.
%FINAL

It is clear that for solar mass primaries in particular, (see niches N1, N3, and N5 in
Table \ref{tab:Niches}) atmospheric mass loss and by proxy water loss,
is expected to be less than that experienced by Earth. This is the
result of early synchronization for periods of up to 50 days depending
upon binary eccentricity. In the case given as niche N3, very little
synchronization takes place. Such niches exist without the aid of
tidal breaking.
%FINAL

\subsection{Multiplanet Habitability}

Most of the examples listed in Table \ref{tab:Niches} have wider
habitable zones than that of the sun, roughly 0.4
AU. These niches have not all been optimized for CHZ width which occurs
for the lowest eccentricity binaries. Solar mass twins, see Niches N1,
N3, and N5, again, provide excellent habitability conditions having
CHZs up to about 50 $\%$ wider than the solar CHZ.
%FINAL

Circumbinary HZs are naturally more distant from the binary center of
mass than those of single stars of the same type as the primary. This
is simply because there are two sources of radiation rather than
one. Having a more distant HZ has an important effect for circumbinary
planets. Because the Hill radius of a circumbinary planet depends
directly on the semi-major axis of the planet and only weakly on the
mass of the binary, circumbinary planets may have moons located
farther from their host planet than is possible for single star HZ
planets. \citet{Heller13b} investigated habitability of exomoons and
find that moons may be adversely affected by the magnetospheres of the
host planet. In addition, moons too close to their planets may undergo
significant tidal heating \citep{Heller13a}. A detailed study of the
habitability of circumbinary exomoons is beyond the scope of the
current paper, but larger Hill radii allow moons to be located farther
from their host planet and thereby they may avoid harmful effects of planetary
aggression. For an interesting discussion on the potential habitability 
of exomoons at the Hill radius see \citet{Hinkel13}. 
%FINAL

\subsection{Long Habitability Lifetimes}

A clear benefit is gained if life may be supported by lower mass
stars.  For circumbinary planets, the lifetime of habitability is
limited by the evolutionary lifetime of the primary star. As one would
expect, the continuous habitable zone lifetime
increases for lower mass primaries. Ultimately for long lived stars,
it is the geothermal lifetime of the planet that determines it's habitability
 lifetime, barring planetary scale
intelligent engineering to prolong planetary habitability.
%FINAL

The points that follow apply equally to single stars and they do to
binaries, because they concern the relation between the narrowing of
the CHZ in comparison to the instantaneous HZ. In each of the examples
shown in Figure \ref{fig:CHZ-Niches}, BHM operates, providing
reduced XUV flux, especially at early times, and lower atmosphere mass
loss than Earth has experienced for many billions of years. However,
there are a few salient points to consider.  In the top two panels of
Figure \ref{fig:CHZ-Niches}, the CHZ of binaries with 1.0 $M_{\odot}$
primaries are shown. We choose a 12.5 Gyr lifetime (of order the age of
the galaxy) as longer lifetimes are likely limited by planetary
factors.  For solar-twins, the CHZ is less than half the width
of the instantaneous CHZ. This effect is seen for solar-like HZ in
Figure \ref{fig:SolarCHZ}, where it is apparent that the Earth is not
in the CHZ of the solar system. In the bottom two panels of Figure
\ref{fig:CHZ-Niches}, the CHZ of binaries with 0.85 $M_{\odot}$
primaries are shown.  Over the course of 12.5 Gyr, the instantaneous HZ
moves much more slowly outward for 0.85 $M_{\odot}$
primaries. For lower mass primaries the CHZ is essentially the same
width as the instantaneous HZ.  Hence, it is more likely for a planet
in the instantaneous HZ to be in the CHZ if it has a lower mass
primary.
%FINAL

\subsection{Photosynthetic Photon Flux Density}

\begin{figure*}
\centering
\includegraphics[width=0.45\textwidth]{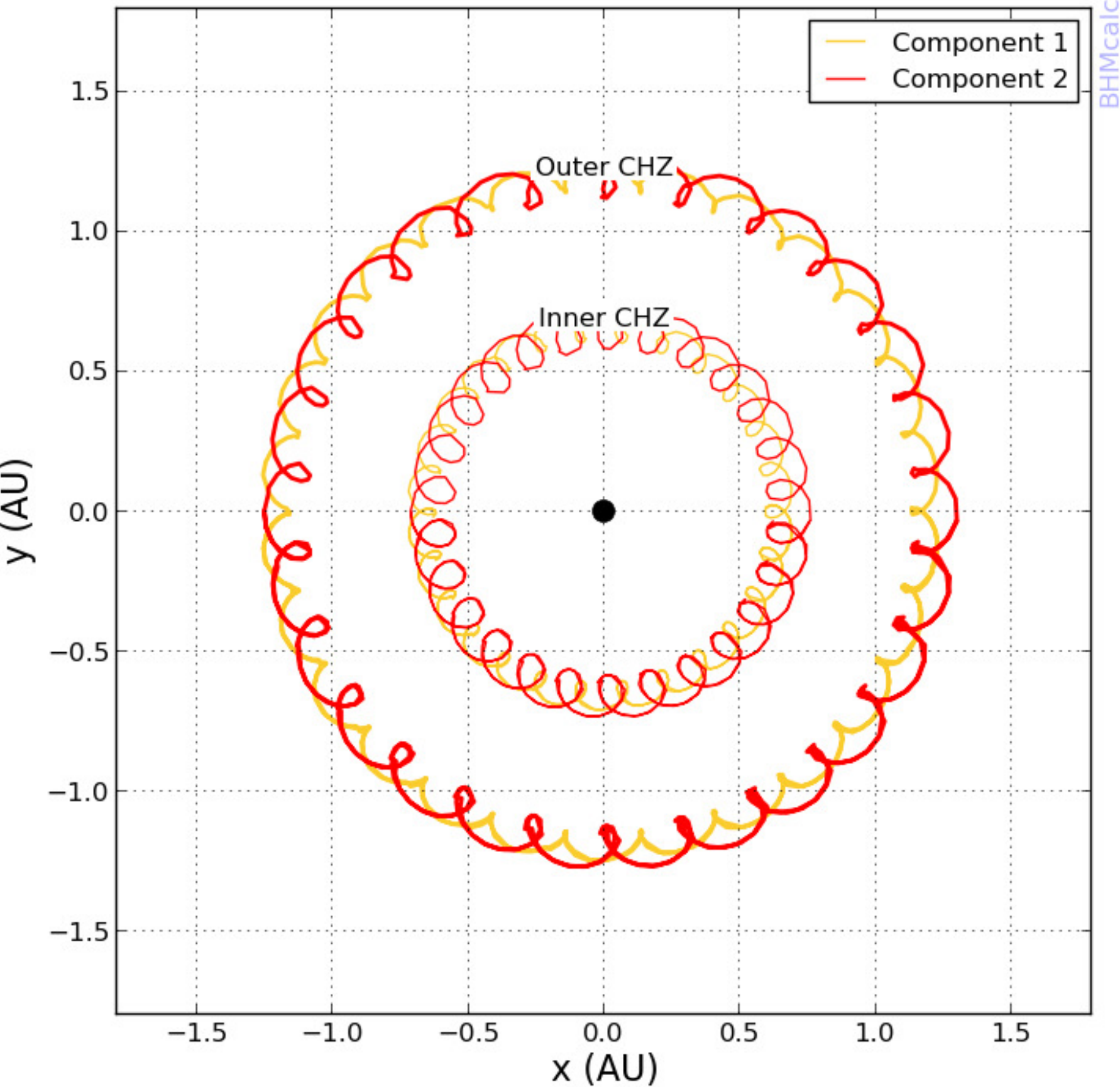}
\hspace{1mm}
\includegraphics[width=0.45\textwidth]{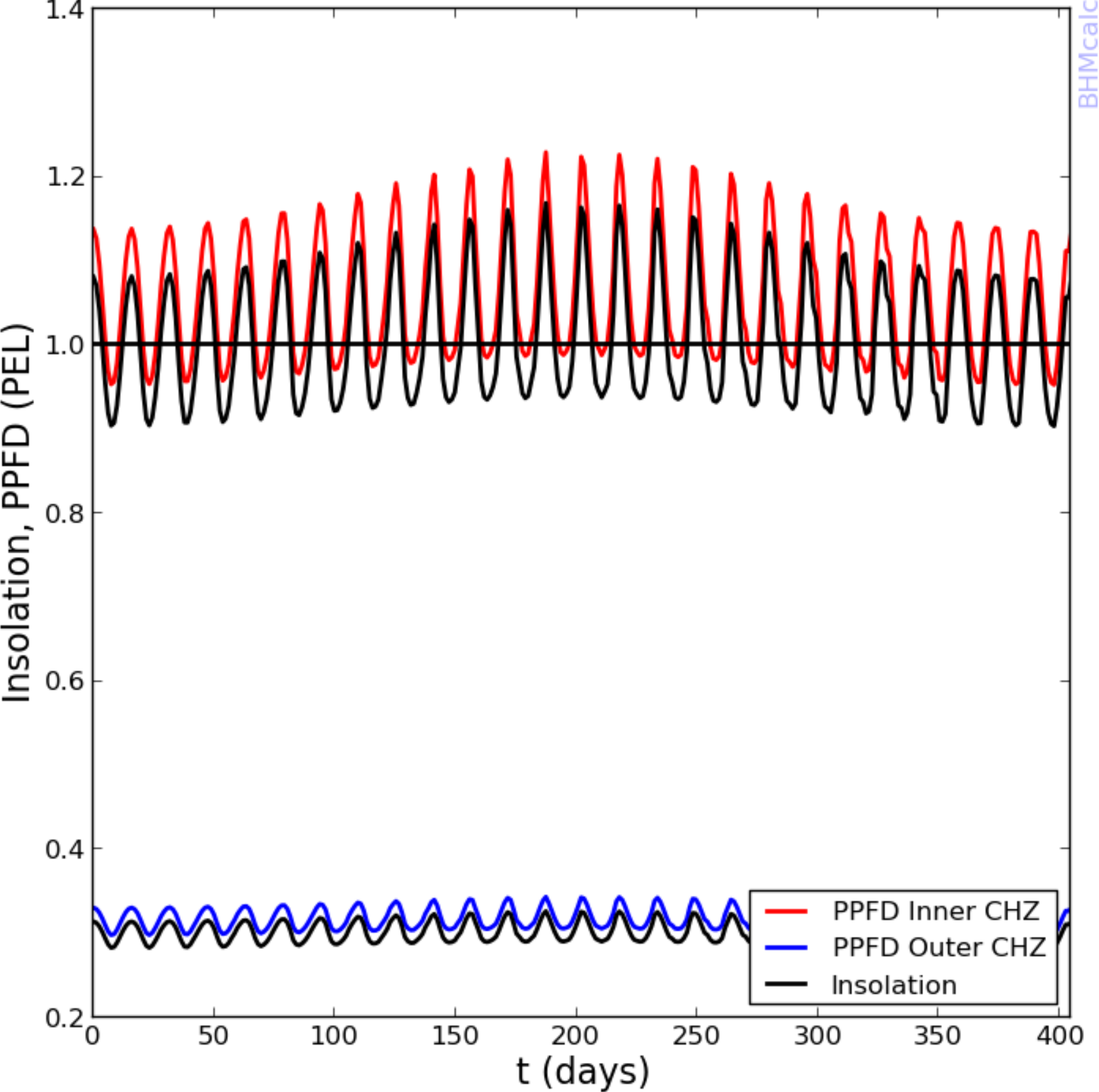}
\caption{In the left panel, planetocentric coordinates (in AU) are used to
  show changes in distance between the planet and the two stars of
  niche N9. The primary is shown in yellow and the secondary is shown
  in red. Two planets, with circular orbits, at the inner and outer CHZ limits
  are shown. In the right panel, the insolation (black) and the PPFD (red for the inner
  planet and blue for the outer planet) are shown in Earth equivalent units.  
  The PPFD curve is higher than the insolation curve for both planets. 
  Specifically, notice that the inner CHZ planet has an insolation about 
  the same as Earth while the PPFD is significantly higher.
\label{fig:PPFD}\vspace{0.4cm}}
\end{figure*}

Photosynthetic photon flux density (hereafter PPFD) is defined as the
number of photons incident per unit
time on a unit surface area at the top of the atmosphere within
some wavelength range considered to be available for use 
by photosynthetic life. See \citet{Kiang07a, Kiang07b} for detailed 
studies of this topic. Here we consider a PPFD bandpass of 4000-14,000\AA \
as a compromise between the extreme ranges used in those papers. 
In Figure \ref{fig:PPFD}, planet centered coordinates, stellar insolation, 
as well as the PPFD received by a circumbinary planet
orbiting in niche 9 is shown. From these plots 
we can see that variation in flux is increased for planets close to the binary. 
Variation in flux decreases farther from the binary, while insolation received
 near the outer edge of the CHZ is at a minimum level. 
 The examples shown are for planets with zero eccentricity. 
 For other cases, we refer the reader to the BHM calculator 
 where these conditions may be explored. As long as the planet 
 remains within the CHZ at all times, variability of insolation and 
 PPFD will remain within the limits given by the examples shown.
     
Both insolation and PPFD are calculated at the sub-solar point (in the
reference case of the Earth) and at the sub-binary center of mass in
the circumbinary case. The time-averaged insolation of the
circumbinary planet located at $CHZ_{in}$ is nearly equal to that of Earth. 
As seen from Figure \ref{fig:PPFD}, the PPFD incident on the circumbinary
planet at the inner CHZ edge is a factor of $\sim$ 1.1 times that incident on
Earth. The implication is that the additional PPFD provided by the
lower mass companion to a solar mass primary will supply increased
energy to the biosphere over that available for Earth life. That is,
a perfect Earth analog of a binary of this type might be expected to 
generate significantly more biomass than the Earth.

%FFFFFFFFFFFFFFFFFFFFFFFFFFFFFFFFFFFFFFFFFFFFFFFFFFFFFFFFFFFFFFFFFFFFFFFFFFFFFFFF
%FIGURE 7
%FFFFFFFFFFFFFFFFFFFFFFFFFFFFFFFFFFFFFFFFFFFFFFFFFFFFFFFFFFFFFFFFFFFFFFFFFFFFFFFF
%FFFFFFFFFFFFFFFFFFFFFFFFFFFFFFFFFFFFFFFFFFFFFFFFFFFFFFFFFFFFFFFFFFFFFFFFFFFFFFFF% 

%%%%%%%%%%%%%%%%%%%%%%%%%%%%%%%%%%%%%%%%%%%%%%%%%%%%%%%%%%%%%%%%%%%%%%%%%
\section{Discussion and Conclusions}
\label{sec:Conclusion}
%%%%%%%%%%%%%%%%%%%%%%%%%%%%%%%%%%%%%%%%%%%%%%%%%%%%%%%%%%%%%%%%%%%%%%%%%

In order to properly address the possibility of life on exoplanets,
factors beyond the standard concept of the HZ must be considered.
\citet{Mason13} identified a novel mechanism potentially enhancing
habitability of circumbinary planets with respect to those found
around single stars. In summary, the early tidal breaking of the
primary's rotation results in a reduction of XUV and SW flux, which
is expected to cause mass loss from planetary atmospheres.  
Main sequence binaries
with periods in the 10-50 day range often provide excellent habitable
environments, within which life may thrive.  Planets and moons in
these HZs need less magnetic protection than their single star
counterparts.

We have shown that BHM allows superlative circumbinary HZs.  One
of the implications of this discovery is that many binaries could
potentially harbor planetary systems with more than one habitable
planet per system increasing our chances to find habitable planets in
the Galaxy. To see why this could be propitious for life, consider
that many constraints on life are specific to entire planetary
systems, independent of the number of worlds in habitable zones. For
instance, life on rocky planets or moons likely requires the right
abundance of volatiles and radiogenic elements for prolonged geologic
activity. Catastrophic sterilization events such as nearby supernovae
and gamma-ray bursts affect entire planetary systems not just specific
worlds. Giant planets may either enhance or disrupt the development of
complex life within a given system, irrespective of how many habitable
planets exist in the HZ.

It might be rare for a particular planetary system to possess
qualities that promote life and to be lucky enough to avoid
cataclysm. However, multiple habitable planets within a planetary
system may possess improved chances for advanced life to develop as
panspermia may occur in planetary systems with several habitable
worlds. We speculate that exomoons within a BHM protected HZ,
may be habitable in some cases as a larger Hill radius allows
more distant exomoon orbits. The best predictor of life on one HZ
planet might be the presence of life on its neighbor. Reduced stellar
aggression, as the result of BHM, may mean that circumbinary
habitability goes hand in hand with planetary systems with multiple
habitable worlds.

We find that habitability around binary stars is more complex than
that of single stars. Due to this added complexity, especially via the
operation of BHM; there is a broad diversity in conditions
experienced by circumbinary planets. Single star habitability on the
other hand, appears to be a trade-off between long lifetimes and
prolonged desiccative activity of low mass stars, versus the high
luminosity, but short lifetimes of high mass main sequence stars. The
sun appears to lie somewhere within a fairly narrow single star
habitable niche.  However, in BHM binaries that have lower than solar mass
primaries, long stellar lifetimes are accompanied by reduced
activity. Obvious advantages of long lifetimes include more time for
the origin and development of life and more time for life to recover
from catastrophes producing mass extinctions.

In addition, the combined spectrum incident 
on some circumbinary planets may
allow for the more rapid recovery from global catastrophes, because,
in some cases circumbinary planets are bathed in considerable
photosynthetic flux. Such high flux may allow for quick recovery, due
to enhanced biomass production. Such cataclysms include impacts,
supernovae and gamma-ray burst sterilization, and hyper-volcanism. 
It is possible that
life is capable of modulating planetary conditions such as
atmospheric composition, soil content, and temperature. This is the
so-called Gaia effect \citep{Lovelock72a}. Namely, life forms of a
planet may adapt so well that they eventually couple with their
environment in such a way that they act as a single self-regulating
system. It is not clear whether or not the Gaia effect works on
Earth. However, It is natural to expect that the effectiveness of life
to regulate habitability conditions of its host planet might be a
direct function of the biomass the planet can manage to produce
\citep{Zuluaga14}. Photosynthetic photon flux density in excess of
Earth levels suggests that these circumbinary planets should be
considered to be super habitable. 

Upcoming missions to search for planets such as the Transiting Exoplanet
Survey Satellite (\textit{TESS}) \citep{Ricker10} and the Planetary Transits 
and Oscillations of Stars Telescope (\textit{PLATO})
\citep{Rauer13} are expected to discover many new planets. Some of
these planets will be orbiting binaries. TESS is not 
optimized for planets located in the HZ of their host stars as they are most
sensitive to short period planets, since it will observe most 
fields for only $\sim $27 days. Nevertheless, it will likely find many small planets 
located in or near the solar neighborhood. Follow-up studies of these 
nearby putative planetary systems may reveal additional planets with larger 
orbits. PLATO is expected to discover many more planets and is sensitive
to planets orbiting in the HZ. Detection of circumbinary planets is intrinsically 
more difficult than finding planets around single stars. However, the success of the 
Kepler mission in finding circumbinary planets despite no expectations, is encouraging. 
See \citet{Welsh14} for a review of these discoveries.

Circumbinary HZ planets have been discovered and a mechanism for 
sustained habitability in these systems has been established. 
We therefore strongly suggest that binaries with BHM protected
HZs are strong candidates for extraterrestrial life and should be
included in programs to detect habitable planets.

%FINAL

% %%%%%%%%%%%%%%%%%%%%%%%%%%%%%%%%%%%%%%%%%%%%%%%%%%%%%%%%%%%%%%%%%%%%%%
% \section*{Acknowledgments}
% %%%%%%%%%%%%%%%%%%%%%%%%%%%%%%%%%%%%%%%%%%%%%%%%%%%%%%%%%%%%%%%%%%%%%%

\acknowledgments

This work is supported in part from incentive funds to P. A. Mason
associated with an NSF/PAARE grant to the University of Texas at El
Paso. J.I. Zuluaga and P.A. Cuartas-Restrepo are supported by
CODI-UdeA.

%%%%%%%%%%%%%%%%%%%%%%%%%%%%%%%%%%%%%%%%%%%%%%%%%%%%%%%%%%%%%%%%%%%%%%%%%%%%%%%%%
%BIBLIOGRAPHY
%%%%%%%%%%%%%%%%%%%%%%%%%%%%%%%%%%%%%%%%%%%%%%%%%%%%%%%%%%%%%%%%%%%%%%%%%%%%%%%%%

\bibliography{bibliography}
\bibliographystyle{ApJ}

%%%%%%%%%%%%%%%%%%%%%%%%%%%%%%%%%%%%%%%%%%%%%%%%%%%%%%%%%%%%%%%%%%%%%%%%%%%%%%%%%
%FIGURES
%%%%%%%%%%%%%%%%%%%%%%%%%%%%%%%%%%%%%%%%%%%%%%%%%%%%%%%%%%%%%%%%%%%%%%%%%%%%%%%%%

\end{document}